\documentclass[draftclsnofoot,onecolumn,12pt]{IEEEtran}
\usepackage{amsfonts}
\usepackage{amsmath}
\usepackage{amssymb}
\usepackage{graphicx}
\providecommand{\U}[1]{\protect\rule{.1in}{.1in}}
\newtheorem{theorem}{Theorem}

\newtheorem{conjecture}[theorem]{Conjecture}
\newtheorem{corollary}[theorem]{Corollary}

\newtheorem{definition}[theorem]{Definition}
\newtheorem{example}[theorem]{Example}

\newtheorem{lemma}[theorem]{Lemma}

\newtheorem{proposition}[theorem]{Proposition}
\newtheorem{remark}{Remark}

\usepackage{graphicx,epsf,psfrag}
\usepackage{amsmath,amssymb,amsfonts,verbatim}
\usepackage{mathrsfs}
\usepackage{etoolbox}

\usepackage{latexsym}

\newcommand{\beq}{\begin{equation}}
\newcommand{\eeq}{\end{equation}}
\newcommand{\eps}{\epsilon}

\newcommand{\bi}{\begin{itemize}}
\newcommand{\ei}{\end{itemize}}

\newcommand{\calA}{\mathcal{A}}

\newcommand{\calE}{\mathcal{E}}

\newcommand{\calP}{\mathcal{P}}

\newcommand{\calX}{\mathcal{X}}

\newcommand{\bu}{\mathbf{u}}

\newcommand{\bbE}{\mathbb{E}}

\newcommand{\bbN}{\mathbb{N}}

\newcommand{\bbP}{\mathbb{P}}

\newcommand{\bbR}{\mathbb{R}}

\newcommand{\bbZ}{\mathbb{Z}}

\DeclareMathAlphabet{\mathbsf}{OT1}{cmss}{bx}{n}
\DeclareMathAlphabet{\mathssf}{OT1}{cmss}{m}{sl}

\DeclareSymbolFont{bsfletters}{OT1}{cmss}{bx}{n}  
\DeclareSymbolFont{ssfletters}{OT1}{cmss}{m}{n}
\DeclareMathSymbol{\bsfGamma}{0}{bsfletters}{'000}
\DeclareMathSymbol{\ssfGamma}{0}{ssfletters}{'000}
\DeclareMathSymbol{\bsfDelta}{0}{bsfletters}{'001}
\DeclareMathSymbol{\ssfDelta}{0}{ssfletters}{'001}
\DeclareMathSymbol{\bsfTheta}{0}{bsfletters}{'002}
\DeclareMathSymbol{\ssfTheta}{0}{ssfletters}{'002}
\DeclareMathSymbol{\bsfLambda}{0}{bsfletters}{'003}
\DeclareMathSymbol{\ssfLambda}{0}{ssfletters}{'003}
\DeclareMathSymbol{\bsfXi}{0}{bsfletters}{'004}
\DeclareMathSymbol{\ssfXi}{0}{ssfletters}{'004}
\DeclareMathSymbol{\bsfPi}{0}{bsfletters}{'005}
\DeclareMathSymbol{\ssfPi}{0}{ssfletters}{'005}
\DeclareMathSymbol{\bsfSigma}{0}{bsfletters}{'006}
\DeclareMathSymbol{\ssfSigma}{0}{ssfletters}{'006}
\DeclareMathSymbol{\bsfUpsilon}{0}{bsfletters}{'007}
\DeclareMathSymbol{\ssfUpsilon}{0}{ssfletters}{'007}
\DeclareMathSymbol{\bsfPhi}{0}{bsfletters}{'010}
\DeclareMathSymbol{\ssfPhi}{0}{ssfletters}{'010}
\DeclareMathSymbol{\bsfPsi}{0}{bsfletters}{'011}
\DeclareMathSymbol{\ssfPsi}{0}{ssfletters}{'011}
\DeclareMathSymbol{\bsfOmega}{0}{bsfletters}{'012}
\DeclareMathSymbol{\ssfOmega}{0}{ssfletters}{'012}

\newcommand{\bari}{\bar{i}}

\newcommand{\barP}{\bar{P}}

\newcommand{\balpha}{\bm{\alpha}}

\newcommand{\floor}[1]{\lfloor{#1}\rfloor}

\DeclareMathOperator*{\argmax}{arg\,max}
\DeclareMathOperator*{\argmin}{arg\,min}

\DeclareMathOperator{\cov}{Cov}

\ifcsmacro{theorem}{}{
\newtheorem{theorem}{Theorem}
\newtheorem{lemma}[theorem]{Lemma}
\newtheorem{proposition}[theorem]{Proposition}
\newtheorem{corollary}[theorem]{Corollary}

\newtheorem{remark}{Remark}

}

\newcommand{\qednew}{\nobreak \ifvmode \relax \else
      \ifdim\lastskip<1.5em \hskip-\lastskip
      \hskip1.5em plus0em minus0.5em \fi \nobreak
      \vrule height0.75em width0.5em depth0.25em\fi}

\begin{document}

\title{Asymptotics and Non-asymptotics for Universal Fixed-to-Variable Source Coding}
\author{Oliver Kosut, \IEEEmembership{Member, IEEE} and Lalitha Sankar, \IEEEmembership{Member, IEEE}
\thanks{O. Kosut and L. Sankar are with the School of Electrical, Computer and Energy Engineering, Arizona State University (Email:  okosut@asu.edu, lalithasankar@asu.edu).   } \thanks{This paper was presented in part at the International Symposia on Information Theory in 2013 \cite{Kosut2013} and 2014 \cite{Kosut2014}.}}
\maketitle

\begin{abstract}
Universal fixed-to-variable lossless source coding for memoryless sources is studied in the finite blocklength and higher-order asymptotics regimes. Optimal third-order coding rates are derived for general fixed-to-variable codes and for prefix codes. It is shown that the non-prefix \emph{Type Size code}, in which codeword lengths are chosen in ascending order of type class size, achieves the optimal third-order rate and outperforms classical Two-Stage codes. Converse results are proved making use of a result on the distribution of the empirical entropy and Laplace's approximation. Finally, the fixed-to-variable coding problem without a prefix constraint is shown to be essentially the same as the universal guessing problem.
\end{abstract}

\section{Introduction}

We have entered an era in which large volumes of data are continually
generated, accessed, and stored across distributed servers. In contrast to the
traditional data communications models in which large blocks of data are
compressed, the evolving information generation, access, and storage contexts
require compressing relatively smaller blocks of data asynchronously and
concurrently from a large number of sources. Typical examples include online
retailers and social network sites that are continuously collecting, storing,
and analyzing user data for a variety of purposes. Finite blocklength
compression schemes could be well suited to these applications.

The finite blocklength (near-) lossless source coding literature typically
assumes knowledge of the underlying source distribution
\cite{VerduYannis,YannisVerdu}. In general, however, the distribution may
neither be known \textit{a priori }nor easy to estimate reliably in the small
blocklength regime. The cost of universality in lossless source coding has
been studied in \cite{Clarke1990}, and more recently in the finite length
regime by \cite{Beirami2011a}. In contrast to these works, our work does not
assume that a prefix-free code; furthermore, in place of redundancy, our
performance metric bounds the probability of the code-length exceeding a given
number of bits, which we call the \emph{$\eps$-rate}, as it is more in-keeping with the finite blocklength literature (e.g.,
\cite{VerduYannis,YannisVerdu}). These appear to change the problem, as our achievable and
converse bounds on the third-order coding rate differs (are tighter) from the
corresponding one from \cite{Clarke1990}. More recently, \cite{Beirami2014} proved a general converse for universal prefix-free coding of parametric sources under the redundancy metric, and found similar results to ours.

We consider fixed-to-variable length coding schemes for a stationary
memoryless---often referred to as independent and identically distributed
(i.i.d.)---source with unknown distribution $P$. For such a source, the
minimal rate required to compress a length $n$ sequence with
probability $(1-\epsilon)$ is given by\footnote{$Q$ is the Gaussian cdf
$Q(x)=\frac{1}{\sqrt{2\pi}}\int_{x}^{\infty}e^{-t^{2}/2}dt$, and $Q^{-1}$ is
its inverse function. $H(P)$ and $V(P)$ are the entropy and varentropy respectively of distribution $P$. See Sec.~\ref{Sec_Prelim} for formal definitions.}
\begin{equation}
H(P)  +\sqrt{\frac{V(P)}{n}}Q^{-1}(\epsilon)+c\frac{\log(n)}{n}+O\left(\frac{1}{n}\right).
\label{Rstar}%
\end{equation}
The first term is the usual entropy, giving the best asymptotically achievable rate. The second term is the so-called dispersion, characterizing the additional required data rate due to random variation in the information content of the source sequence. The third term is the main focus of this work, as it is the largest term in which the cost of universality is evident. When the source distribution is known \cite{Strassen}%
\footnote{Although there is a gap in Strassen's original proof; see discussion
following (129) in \cite{YannisVerdu}.}, the third-order coefficient is given by $c=-1/2$; it was further pointed out in \cite{YannisVerdu} that this is the optimal third-order rate whether or not the prefix code restriction is in place. We find that in the universal setting, the optimal third-order coefficient becomes
\beq\label{eq:3rd_order}
c=\frac{|\calX_P|-3}{2}
\eeq
where $\calX_P$ is the support set of the source distribution $P$. Achievability is proved using the Type Size code, wherein sequences are coded in increasing order of type class size. This code differs from the Two-Stage code, a common approach to fixed-to-variable universal coding in which the type of the source sequence is encoded, followed by the index of he sequence within its type class \cite[Chap. 13, pp. 433]{Cover:book}. We find that the Type Size codes outperforms Two-Stage codes in third-order coding rate. Subsequent to our introduction of the Type Size code in \cite{Kosut2013}, \cite{Beirami2014} shows that the Type Size code is minimax optimal with respect to redundancy. To prove that \eqref{eq:3rd_order} is the optimal, we prove a converse using a characterization of the distribution of the empirical entropy, as well as an application of Laplace's approximation.

While the above results apply for codes that are not restricted to be prefix codes, we also find that subject to this restriction, the optimal third-order coefficient is
\beq\label{eq:3rd_order_prefix}
c=\frac{|\calX_P|-1}{2}.
\eeq
While the difference in coding rates between \eqref{eq:3rd_order} and \eqref{eq:3rd_order_prefix}
 may seem small, when the compression algorithm is
used very many times over small blocks of data, this difference can
significantly affect storage capability. An example of such a use is storage
in social networks wherein updates of every user are asynchronously
compressed as they arrive resulting cumulatively in an extremely large
number of uses of the compression algorithm.

\begin{table}\label{summary}
\begin{center}
\begin{tabular}{r | l | l | l | l |}
\cline{2-5}
&\multicolumn{2}{c|}{$\eps$-Rate Third Order Term} & \multicolumn{2}{c|}{Redundancy}\\ \cline{2-5}
& \multicolumn{1}{c|}{Non-prefix} & \multicolumn{1}{c|}{Prefix} & \multicolumn{1}{c|}{Non-prefix} & \multicolumn{1}{c|}{Prefix} \\ \hline
\multicolumn{1}{|c|}{Non-universal}& $-\frac{1}{2}\frac{\log n}{n}$ \cite{YannisVerdu} & $-\frac{1}{2}\frac{\log n}{n}$ \cite{YannisVerdu}
& $-\frac{1}{2}\frac{\log n}{n}$ \cite{Szpankowski2011} & $0$ \cite{Shannon1948} \\ \hline
\multicolumn{1}{|c|}{Universal} & $\frac{d-2}{2}\frac{\log n}{n}$ [present] & $\frac{d}{2}\frac{\log n}{n}$ [present]
& $\frac{d-2}{2}\frac{\log n}{n}$ \cite{Beirami2014} & $\frac{d}{2}\frac{\log n}{n}$ \cite{Clarke1990} \\ \hline
\multicolumn{1}{|c|}{Difference} & $\frac{d-1}{2}\frac{\log n}{n}$ & $\frac{d+1}{2}\frac{\log n}{n}$ 
& $\frac{d-1}{2}\frac{\log n}{n}$ & $\frac{d}{2} \frac{\log n}{n}$ \\ \hline
\end{tabular}
\end{center}
\caption{Rate results for non-universal and universal prefix and non-prefix codes, measuring both $\eps$-rate and redundancy, excluding  $O(\frac{1}{n})$ terms. For $\eps$-rate only the third-order term is given (the first two can be seen in \eqref{Rstar}), and the redundancy is normalized by the blocklength $n$ for comparison. The third row gives the difference between the non-universal and universal rates (i.e. the cost of universality). All rates are given in terms of the dimension of the space of distributions, which for i.i.d. distributions is $d=|\calX|-1$. Citations are given in which each result is proved (those for the $\eps$-rate of universal codes are proved in the present paper).}
\end{table}

Our results are summarized and compared to prior findings in Table~\ref{summary}, where the results are given in terms of the dimension $d$ of the set of possible distributions. In our case, we consider all i.i.d. distributions on the alphabet $\calX$, so the dimension is that of the simplex, i.e. $d=|\calX|-1$. Shown in Table~\ref{summary} are the relevant rate terms both in terms of $\eps$-rate and redundancy. Our results are along the lines of \cite{Clarke1990}, which found that for a parametric source with dimension $d$, the best achievable redundancy of a universal prefix code is roughly $\frac{d}{2}\frac{\log n}{n}$. As $\eps$-rate is a more refined metric than redundancy, our results can be used to recover those of \cite{Clarke1990} in the case of i.i.d. sources (although their results were more general).

From Table~\ref{summary}, one can see that the difference in rate for optimal non-prefix and prefix universal codes is roughly $\frac{\log n}{n}$. Two effects account for this difference, each contributing $\frac{1}{2}\frac{\log n}{n}$:
\begin{enumerate}
\item There is a difference of $\frac{1}{2}\frac{\log n}{n}$ between non-prefix and prefix even in the non-universal setting. This difference appears in Table~\ref{summary} for redundancy, but not for $\eps$-rate. This is because, even though, as proved in \cite{YannisVerdu}, both non-prefix and prefix codes can achieve a third-order rate of $-\frac{1}{2}\frac{\log n}{n}$, the non-prefix code does not depend on $\eps$, while the prefix code does.\footnote{The prefix code in \cite{YannisVerdu} is a two-level code, assigning the most likely sequences a short length, and the less likely sequences a long length.} To achieve universality in $\eps$ costs $\frac{1}{2}\frac{\log n}{n}$ in rate.
\item Without a prefix constraint, codewords of different lengths do not affect one another: they do not compete for `codeword space'. Thus the additional rate needed for universality depends only on the dimension of the manifold of distributions with roughly the same entropy, which is $|\calX|-2$ or $d-1$. This leads to a third-order rate $\frac{d-1}{2}\frac{\log n}{n}$ larger than the non-universal rate. With the prefix constraint, codewords of different lengths do affect one another, so the relevant dimension is $|\calX|-1$ or $d$, leading to a third-order rate $\frac{d}{2}\frac{\log n}{n}$ larger.
\end{enumerate}

The paper is organized as follows. In Sec.~\ref{Sec_Prelim}, we introduce
the finite-length lossless source coding problem, performance metrics, and
related definitions. In Sec.~\ref{Guessing}, we relate the fixed-to-variable coding problem without a prefix constraint to the universal guessing problem, in which a source sequence is successively guessed until correctly identified, and we show that these two problems are essentially the same. In Sec.~\ref{SecBinary}, we explore in detail the case of binary sources. In Sec.~\ref{SecPrelim}, we provide several preliminary results to be used in our main achievability and converse proofs. These include a precise characterization of the distribution of the empirical entropy, as well as an exploration of type class size. In Sec.~\ref{SecAch}, we present results on specific achievable schemes, namely Two-Stage codes and the Type Size code. In Sec.~\ref{SecConv}, we present our converse results, both for general fixed-to-variable codes and those restricted to be prefix codes.  We conclude in Section \ref{SecConc}.

\section{\label{Sec_Prelim}Problem Setup}

First, a word on the nomenclature in the paper: we use $\mathbb{P}$ to denote
probability with respect to distribution $P$, $\bar{\mathbb{P}}$ for
probability with respect to distribution $\bar{P}$, and $\mathbb{E}$ to
denote expectation. All logarithms are with respect to base 2. Let
$\mathcal{P}$ be the simplex of distributions over the finite alphabet
$\mathcal{X}$. Given a distribution $P\in\calP$, $\calX_P$ denotes the support set of $P$; i.e., $\calX_P=\{x\in\calX:P(x)>0\}$. Define the information under a distribution $P$ as
\begin{equation}
\imath_{P}(x):=\log\frac{1}{P(x)}. \label{InfoRV}%
\end{equation}
The source entropy and the varentropy are given as $H(P):=\mathbb{E}[\imath_{P}(X)]$ and $V(P):=\text{Var}[\imath_{P}(X)]$
where the expectation and variance are over $P.$ We will sometimes abbreviate these as $H$ and $V$, when the distribution $P$ is clear form context. Given a sequence $x^{n}\in\mathcal{X}^{n}$, let $t_{x^{n}}$ be the type of
$x^{n}$, so that
\begin{equation}
t_{x^{n}}(x):=\frac{|\{i:x_{i}=x\}|}{n}.
\end{equation}
For a type $t$, let $T_{t}$ be the type class of $t$, i.e.
\begin{equation}
T_{t}:=\{x^{n}\in\mathcal{X}^{n}:t_{x^{n}}=t\}.
\end{equation}

We consider a universal source coding problem in which a single code must
compress a sequence $X^{n}$ that is the output of an i.i.d. source with single
letter distribution $P$, where $P$ may be any element of $\mathcal{P}%
$. Any $n$-length sequence from the source is coded to a variable-length bit
string via a coding function
\begin{equation}
\phi:\mathcal{X}^{n}\rightarrow\{0,1\}^{\star}=\{\emptyset
,0,1,00,01,10,11,000,\ldots\}.
\end{equation}
A \emph{prefix code} $\phi$ is one such that for any pair of sequences $x^n,x'^n\in\calX^n$, $\phi(x^n)$ is not a prefix of $\phi(x'^n)$. In general, we do not restrict only to prefix codes, but some results will apply for this subclass. Let $\ell(\phi(x^{n}))$ be the number of bits in the compressed binary string when
$x^{n}$ is the source sequence. The figure of merit is the $\epsilon$-coding rate $R(\phi;\epsilon,P)$, the minimum rate such that the probability of exceeding it is at most $\epsilon$; that is
\begin{equation}\label{Rate_def}
R(\phi;\epsilon,P)=\min\left\{ \frac{k}{n}:\mathbb{P}(\ell(\phi(X^n))>k)\le \epsilon\right\}.
\end{equation}
We say a rate function $R(\eps,P)$ is \emph{$n$-achievable} if there exists an $n$-length fixed-to-variable code $\phi$ satisfying $R(\phi;\eps,P)\le R(\eps,P)$ for all $\eps,P$. Note that this figure of merit is not a single number, or even a finite-length vector: it is a function of the continuous parameters $\eps$ and $P$.

The above definitions differ from \cite{Clarke1990,Beirami2011a} in two ways.
First, they assume prefix-free codes. Second, for the figure of merit they use
\emph{redundancy}, defined as the difference between the expected code length
and the entropy of the true distribution:
\begin{equation}
\mathbb{E}\left[  \ell(\phi(X^{n}))-\log\frac{1}{P^n(X^{n})}\right]
\end{equation}
where the expectation is taken with respect to $P$. Using $\epsilon
$-coding rate rather than redundancy gives more refined information about the
distribution of code lengths. In \cite{Clarke1990} it is proved that the
optimal redundancy for a universal prefix-free code is given by
$\frac{d}{2}\log n+O(1)$
where $d$ is the dimension of the set of possible source distributions (for
i.i.d. sources, $d=|\mathcal{X}_{P}|-1$ where $\mathcal{X}_{P}$ is the support
of $X$ under $P$). In Sec.~\ref{SecConv}, we show that with our model, there
exists a universal code such that the gap to the optimal number of bits with
known distribution is (using the $d$ notation)
$\frac{d-1}{2}\log n+O(1)$.
Our lack of restriction to prefix-free codes appears to account for the
difference seen between these two results. In Sec. \ref{SecAch}, we show that
for a prefix-free Two-Stage code the gap to the optimal is $\frac{
d+1  }{2}$.

\section{\label{Guessing}Universal Guessing}

The fixed-to-variable coding problem without a prefix constraint is closely related to the so-called guessing problem. First introduced by Massey \cite{Massey1994}, guessing is a variation on source coding in which a random sequence $X^n$ is drawn, and then a guesser asks a series of questions of the form ``Is $X^n$ equal to $x^n$?'' until the answer is ``Yes''. The guesser wishes to minimize the required number of guesses before guessing correctly. In this section we formally describe the universal guessing problem, and demonstrate its relationship to the source coding problem.

We say a function $G:\calX^n\to \{1,\ldots,|\calX|^n\}$ is a \emph{guessing function} if it is one-to-one. Each function $G$ represents a guessing strategy that first guesses $G^{-1}(1)$, then $G^{-1}(2)$, and so forth. Thus $G(x^n)$ is the number of guesses required if $X^n=x^n$. We define the tail probability figure of merit for guessing functions as
\beq
M(G;\eps,P)= \min\{m:\bbP(G(X^n)>m)\le\eps\}.
\eeq
We say $M(\eps,P)$ is \emph{$n$-achievable} if there exists a guessing function $G$ with $n$-length inputs such that $M(G;\eps,P)\le M(\eps,P)$ for all $\eps,P$.

The following theorem relates the set of achievable $R(\eps,P)$ to the set of achievable $M(\eps,P)$. In fact, the theorem asserts that the set of achievable $R(\eps,P)$ is completely determined by the set of achievable $M(\eps,P)$, although not vice versa. Thus, the guessing problem is in some sense strictly more refined that the fixed-to-variable coding problem; still, throughout this paper we present results in terms of the latter, as we believe it to be the more useful problem.

\begin{theorem}
The $\eps$-rate function $R(\eps,P)$ is $n$-achievable if and only if there exists an $n$-achievable $M(\eps,P)$ such that
\beq\label{eq:M_R}
\lfloor\log M(\eps,P)\rfloor \le  n R(\eps,P)\text{ for all }\eps,P.
\eeq
\end{theorem} 
\begin{IEEEproof}
First assume $R(\eps,P)$ is $n$-achievable, and we show that there exists an $n$-achievable $M(\eps,P)$ satisfying \eqref{eq:M_R}. By assumption, there exists an $n$-length code $\phi$ such that $R(\phi;\eps,P)\le R(\eps,P)$. Let $m_k$ be the number of sequences $x^n$ for which $\ell(\phi(x^n))\le k$. We construct a guessing function as follows. For each integer $k$, assign $G(x^n)$ for all $x^n$ for which $\ell(\phi(x^n))=k$ to the integers between $m_{k-1}+1$ and $m_{k}$, in any order. Note that $\bbP(\ell(\phi(X^n))>k)=\bbP(G(X^n)>m_k)$. Let $k=nR(\phi;\eps,P)$ for some $\eps,P$. Hence
\beq
\eps\ge \bbP(\ell(\phi(X^n))>k)=\bbP(G(X^n)>m_k)
\eeq
implying that
\beq
M(G;\eps,P)\le m_k\le 2^{k+1}-1
\eeq
where the last inequality follows because the number of bit strings of length at most $k$ is $2^{k+1}-1$. Therefore $\lfloor\log M(G;\eps,P)\rfloor\le k=nR(\phi;\eps,P)$.

Now we assume that there exists an $n$-achievable $M(\eps,P)$, and we shaw that any $R(\eps,P)$ satisfying \eqref{eq:M_R} is achievable. By assumption, there exists a guessing function $G$ such that $M(G;\eps,P)\le M(\eps,P)$. We construct an $n$-length fixed-to-variable code $\phi$ as follows. For each integer $k$, assign $\phi(x^n)$ to distinct bit strings of length $k$ for each of the $2^k$ sequences $x^n$ for which $2^k\le G(x^n)\le 2^{k+1}-1$. Thus $\ell(\phi(x^n))= \lfloor \log G(x^n)\rfloor$ for all $x^n$, which immediately implies $\floor{\log M(G;\eps,P)}=n R(\phi;\eps,P)$. Therefore any $R(\eps,P)$ satisfying \eqref{eq:M_R} is achievable. 
\end{IEEEproof}

\section{\label{SecBinary}Binary Sources}

We begin by examining universal codes for binary i.i.d. sources. Consider
first the optimal code when the distribution is known. These codes were
studied in detail in \cite{VerduYannis,YannisVerdu}. It is easy to see that
the optimal code simply sorts all sequences in decreasing order of
probability, and then assigns sequences to $\{0,1\}^{\star}$ in this order.
Thus the more likely sequences will be assigned fewer bits. For example,
consider an i.i.d. source with $\mathcal{X}=\{\mathsf{A},\mathsf{B}\}$ where
$P_{X}(\mathsf{A})=\delta$ and $\delta>0.5$. The probability of a sequence is
strictly increasing with the number of $\mathsf{A}$s, so the optimal code will
assign sequences to $\{0,1\}^{\star}$ in an order where sequences with more
$\mathsf{A}$s precede those with fewer. For example, for $n=3$, one optimal
order is (sequences with the same type can always be exchanged)
\begin{equation}
\mathsf{AAA},\mathsf{AAB},\mathsf{ABA},\mathsf{BAA},\mathsf{ABB}%
,\mathsf{BAB},\mathsf{BBA},\mathsf{BBB}.
\end{equation}
Interestingly, this is an optimal code for any binary source with $\delta
\ge0.5$. If $\delta< 0.5$, the optimal code assigns sequences to
$\{0,1\}^{\star}$ in the reverse order. That is, there are only two optimal
codes.\footnote{Here our assumption that the code may not be prefix-free becomes relevant, since it is not the case that there are only two optimal prefix-free codes for binary sources.} To design a universal code, we can simply interleave the beginnings of
each of these codes, so for $n=3$ the sequences would be in the following
order:
\begin{equation}
\mathsf{AAA},\mathsf{BBB},\mathsf{AAB},\mathsf{BBA},\mathsf{ABA}%
,\mathsf{BAB},\mathsf{BAA},\mathsf{ABB}.
\end{equation}
In this order, any given sequence appears in a position at most twice as deep
as in the two optimal codes. Hence, this code requires at most one additional
bit as compared to the optimal code when the distribution is known. This holds
for any $n$, as stated in the following theorem.

\begin{theorem}
Let $R^{\star}(n,\epsilon,P_{X})$ be the optimal fixed-to-variable rate when
the distribution $P_{X}$ is known. If $|\mathcal{X}|=2$, there exists a
universal code achieving
\begin{equation}
nR(n,\epsilon,P_{X})\le nR^{\star}(n,\epsilon,P_{X})+1.
\end{equation}

\end{theorem}

\section{Preliminary Results}\label{SecPrelim}

\subsection{Distribution of the Empirical Entropy}

We begin with a lemma bounding the distribution of the empirical entropy of a
length-$n$ data sequence $X^{n}$. This lemma will be used in both
achievability results as well as converses for both prefix and non-prefix
codes to derive third-order coding rates. The lemma is based on a Proposition
on applying central limit theory for functions of random vectors introduced in
\cite{EbrahimMolavianJazi}. We begin by introducing the proposition first; we have
generalized it to include non-zero mean random random vectors.

\begin{proposition}[\cite{EbrahimMolavianJazi} Prop.~1]
\label{Proposition_ML}Let $\left\{  \mathbf{U}_{t}:=U_{1t},U_{2t}%
,...,U_{Kt}\right\}  _{t=1}^{\infty}$ be i.i.d. random vectors in
$\mathbb{R}^{K}$ with mean $\mathbf{u}_{0}$ and $E\left[  \left\Vert
\mathbf{U}_{1}\right\Vert _{2}^{3}\right]  <\infty,$ and denoting
$\mathbf{u}:=\left(  u_{1},u_{2},\ldots,u_{K}\right)$, let $\mathbf{f}\left(
\mathbf{u}\right)  :\mathbb{R}^{K}\rightarrow\mathbb{R}^{L}$ be an
$L$-component vector function $\mathbf{f}\left(  \mathbf{u}\right)  =\left(
f_{1}\left(  \mathbf{u}\right)  ,f_{2}\left(  \mathbf{u}\right)
,\ldots,f_{L}\left(  \mathbf{u}\right)  \right)  $ which has continuous
second-order partial derivatives in a $K\,$-hypercube neighborhood of
$\mathbf{u}=\mathbf{u}_{0}$ of side length at least $\frac{1}{\sqrt[4]{n}}%
,$and whose corresponding Jacobian matrix $\mathbf{J}$ at $\mathbf{u=u}_{0}$
consists of the following first-order partial derivatives
\begin{equation}%
\begin{array}
[c]{ccc}%
J_{lk}:=\left.  \frac{\partial f_{l}\left(  \mathbf{u}\right)  }{\partial u_{k}%
}\right\vert _{\mathbf{u}=\mathbf{u}_{0}}, & l=1,\ldots,L, & k=1,2,\ldots,K.
\end{array}
\end{equation}
Then, for any convex Borel-measureable set $\mathcal{D}$ in $\mathbb{R}^{L}$,
there exists a finite positive constant $B$ such that
\begin{equation}
\left\vert \bbP\left[  \mathbf{f}\left(  \frac{1}{n}%
{\displaystyle\sum\limits_{t=1}^{n}}
\mathbf{U}_{t}\right)  \in\mathcal{D}\right]  -\bbP\left[  \mathcal{N}\left(
\mathbf{f}\left(  \mathbf{u}_{0}\right)  ,\mathbf{V}\right)  \in
\mathcal{D}\right]  \right\vert \leq\frac{B}{\sqrt{n}},
\end{equation}
where the covariance matrix $\mathbf{V}$\ is given by $\mathbf{V=}\frac{1}%
{n}\mathbf{J}\cov\left(  \mathbf{U}_{1}-\mathbf{u}_{0}\right)  \mathbf{J}^{T},$
that is, its entries are defined as
\begin{equation}%
\begin{array}
[c]{cc}%
V_{ls}:=\frac{1}{n}%
{\displaystyle\sum\limits_{k=1}^{K}}
{\displaystyle\sum\limits_{p=1}^{K}}
J_{lk}J_{sp}\mathbb{E}\left[  (U_{k1}-u_{0k})(U_{p1}-u_{0p})\right]  , & l,s=1,...,L.
\end{array}
\end{equation}

\end{proposition}

\begin{IEEEproof}[Proof sketch] The proof involves three components: (i)\ Taylor
expansion of $\mathbf{f(u})$ about $\mathbf{u}_{0}$ as $\mathbf{f}\left(
\mathbf{u}\right)  =\mathbf{f}\left(  \mathbf{u}_{0}\right)  +\mathbf{J}%
\left(  \mathbf{u-u}_{0}\right)  +\mathbf{R}\left(  \mathbf{u-u}_{0}\right)
$, where the Jacobian matrix $\mathbf{J}$ has entries $J_{kl}=$ $\left.
\frac{\delta f_{k}}{\delta u_{l}}\right\vert _{\mathbf{u}_{0}}$,
and$\ \mathbf{R}\left(  \mathbf{u-u}_{0}\right)  $ is the remainder term which
in the hypercube neighborhood $N\left(  \mathbf{u}_{0},r_{0}\right)  $ of
$\mathbf{u}_{0}$ with side length $r_{0}>\frac{1}{\sqrt[4]{n}},$ can be
bounded by the maximal value of the second order derivatives of $\mathbf{f}%
\left(  \mathbf{u}\right)  $ as
\begin{equation}
\left\vert \mathbf{R}\left(  \mathbf{u}\right)  \right\vert \leq\frac{1}{2}%
\begin{bmatrix}
\displaystyle\max_{1\leq k,p\leq K}\ \max_{\mathbf{u}^{\ast}\in N\left(  \mathbf{u}_{0}%
,r_{0}\right)  }\left\vert \frac{\partial^{2}f_{1}\left(  \mathbf{u}^{\ast
}\right)  }{\partial u_{k}\partial u_{p}}\right\vert \\
\vdots\\
\displaystyle\max_{1\leq k,p\leq K}\ \max_{\mathbf{u}^{\ast}\in N\left(  \mathbf{u}_{0}%
,r_{0}\right)  }\left\vert \frac{\partial^{2}f_{L}\left(  \mathbf{u}^{\ast
}\right)  }{\partial u_{k}\partial u_{p}}\right\vert
\end{bmatrix}
\left(  u_{1}+u_{2}+...+u_{K}\right)  ^{2};
\end{equation}
(ii) bounding the probability that the remainder term concentrates away from
$\mathbf{u}_{0}$ as%
\begin{equation}
\bbP\left[  \left\vert R\left(  \frac{1}{n}%
{\displaystyle\sum\limits_{t=1}^{n}}
\left(  \mathbf{U}_{t}-\mathbf{u}_{0}\right)  \right)  \right\vert >\frac
{1}{\sqrt{n}}\mathbf{1}\right]  \leq\frac{c_{1}}{\sqrt{n}}%
\end{equation}
where we have
\begin{equation}
c_{1}:=\left(  \text{Var}[U_{11}]+...+\text{Var}[U_{K1}]\right)  \left[
1+\frac{K}{2}\min_{1\leq l\leq L}\,\max_{1\leq k,p\leq K}\,\max_{\mathbf{u}^{\ast
}\in N\left(  \mathbf{u}_{0},r_{0}\right)  }\left\vert \frac{\partial^{2}%
f_{l}\left(  \mathbf{u}^{\ast}\right)  }{\partial u_{k}\partial u_{p}}\right\vert
\right]  ;\label{PropML_c1}%
\end{equation}
(iii) bounding
\begin{multline}
\bbP\left[  \mathbf{f}\left(  \frac{1}{n}%
{\displaystyle\sum\limits_{t=1}^{n}}
\mathbf{U}_{t}\right)  \in\mathcal{D}\right]  \leq\label{FinalProb}\\
\bbP\left[  \mathcal{N}\left(  \mathbf{f}\left(  \mathbf{u}_{0}\right)
,\frac{1}{n}\mathbf{J}\cov\left[  \mathbf{U}_{1}\right]  \mathbf{J}^{T}\right)
\in\mathcal{D}\right]  +\frac{c_{3}}{\sqrt{n}}+\frac{c_{2}}{\sqrt{n}}%
+\frac{c_{1}}{\sqrt{n}}%
\end{multline}
where
\begin{equation}
c_{2}=\frac{400L^{1/4}\lambda_{\max}\left(  \mathbf{JJ}^{T}\right)
^{3/2}E\left[  \left\Vert (\mathbf{U}_{1}-\mathbf{u}_0)^T\right\Vert _{2}^{3}\right]
}{\lambda_{\min}\left(  \cov\left[  \mathbf{JU}_1^T\right]  \right)
^{3/2}}%
\end{equation}
and $c_{3}$ results from the Taylor expansion for the probability at hand in a
neighborhood of width $\frac{1}{\sqrt{n}}$ about the set $\mathcal{D}$.
\end{IEEEproof}

\begin{lemma}\label{lemma:empirical_entropy}
Fix positive constant $\beta$ and any distribution $P$ on $\calX$ such that $P(x)\ge\beta$ for all $x\in\calX$ and $V(P)\ge\beta$. Let $X^n\stackrel{\text{i.i.d.}}{\sim} P$. For any $\delta$ and $n$,
\begin{equation}\label{eq:entropy_cdf}
\left\vert \bbP\left[  H\left(  t_{X^{n}}\right)\ge H(P)+\sqrt{\frac{V(P)}{n}}\delta\right]-Q(\delta)\right\vert
 \leq\frac{B}{\sqrt{n}}
\end{equation}
where
\begin{equation}
B=\max\left\{\frac{4}{\beta^2},1+\frac{|\calX|}{\beta}+\frac{400|\calX|^3}{\beta^{3/2}}+\frac{1}{\sqrt{2\pi\beta}}\right\}.
\label{ConstantB}%
\end{equation}
\end{lemma}

\begin{IEEEproof}
We first consider the case that $n\le (2/\beta)^4$. The left hand side of \eqref{eq:entropy_cdf} is at most $1\le \frac{(2/\beta)^2}{\sqrt{n}}\le \frac{B}{\sqrt{n}}$, and we are done.

Now assume $n>(2/\beta)^4$. Let $\mathbf{U}_{i}$ be an $\left\vert \mathcal{X}\right\vert $-length random
vector with entries $U_{i,x}=1\left(  X_{i}=x\right)  $ where $1\left(
\cdot\right)  $ is an indicator function for all $x\in\mathcal{X}$. Note that
$\mathbf{u}_{0}=\mathbb{E}\left[  U_{i,x}\right]  =P\left(  x\right)  ;$
furthermore, $t_{X^{n}}\left(  x\right)  =\frac{1}{n}%
{\textstyle\sum\nolimits_{i=1}^{n}}
U_{i,x}$ and $\cov\left(  \mathbf{U}_{i}\right)  =$ diag$\left\{
\mathbf{P}\right\}  -\mathbf{PP}^{T}$, for all $i,$ where $\mathbf{P}$ is the
vector whose entries are $P\left(  x\right)  $ for all $x\in\mathcal{X}$. 

Let $f(\bu)=\sum_x -u_x\log u_x$ be a scalar function of $\bu$, so that $f\left(  \frac{1}{n}%
{\textstyle\sum\nolimits_{i=1}^{n}}
\mathbf{U}_{i}\right)  =H(t_{X^{n}})$, and let
$\mathcal{D}$ be the half-closed space $[H\left(  P\right)  +\sqrt{\frac
{V(P)}{n}}\delta,\infty).$ Thus, from Applying Proposition
\ref{Proposition_ML}, we have that the left hand side of \eqref{eq:entropy_cdf} is at most $\frac{c_1+c_2+c_2}{\sqrt{n}}$, where the three constants are defined in the proof of Proposition.~\ref{Proposition_ML}. Consider the bound
$c_{1};$ since $\left\vert \partial^{2}f(\mathbf{u})/\partial u_{k}\partial u_{l}\right\vert =$
diag$\left(  1/u_1  ,1/u_1  ,\ldots,1/u_{|\calX|}  \right)  $. Recalling the assumption that $P(x)\ge\beta$ for all $x$, 
\begin{equation}
\max_{%
\mathbf{u}%
\in N(%
\mathbf{u}%
_{0},r_{0})}\left\vert \partial^{2}f/\partial u_{k}\partial u_{l}\right\vert
\le\frac{1}{ \beta-r_0  }.%
\end{equation}
Taking $r_0=n^{-1/4}$, we have that $1/(\beta-r_0)\le 2/\beta$ since $n>(2/\beta)^4$. Hence
\begin{align}
c_{1} &  \leq\sum_{x\in\mathcal{X}}
\left(  P\left(  x\right)  -P^{2}\left(  x\right)  \right) \left(
1+\frac{\left\vert \mathcal{X}\right\vert }{\beta}\right)  \\
&  \leq  1+\frac{\left\vert \mathcal{X}\right\vert }{\beta}.
\label{c1_bound_univ}
\end{align}
The second constant $c_{2}$ can be bounded as%
\begin{equation}
c_{2}\leq\frac{400L^{1/4}\lambda_{\max}\left(  \mathbf{JJ}^{T}\right)
^{3/2}E\left[  \left\Vert (\mathbf{U}_{1}-\mathbf{u}_0)^{T}\right\Vert _{2}^{3}\right]
}{\lambda_{\min}\left(  \cov\left[  \mathbf{JU}_{1}^{T}\right]  \right)
^{3/2}}%
\end{equation}
where $\mathbf{J}=%
\begin{bmatrix}
-1-\log P\left(  1\right)   & -1-\log P\left(  2\right)   & \cdots & -1-\log
P\left(  \left\vert \mathcal{X}\right\vert \right)
\end{bmatrix}
$ such that $\lambda_{\max}\left(  \mathbf{JJ}^{T}\right) =%
{\textstyle\sum\nolimits_{x\in\mathcal{X}}}
\left(  1+\log P\left(  x\right)  \right)  ^{2}\leq\left\vert \mathcal{X}%
\right\vert .$ One can similarly bound $E\left[  \left\Vert (\mathbf{U}_{1}-\mathbf{u}_0)%
^{T}\right\Vert _{2}^{3}\right]  \leq\left\vert \mathcal{X}\right\vert ^{3/2}$
by noting that $\left\Vert (\mathbf{U}_{1}-\mathbf{u}_0)^{T}\right\Vert _{2}\leq
\sqrt{\left\vert \mathcal{X}\right\vert }.$ The term $\cov\left[
\mathbf{JU}_{1}^{T}\right]  =$ Var$\left(
{\textstyle\sum_{x\in\mathcal{X}}}
\left(  -1-\log P\left(  x\right)  \right)  U_{1,x}\right)  =$ Var$\left(
{\textstyle\sum_{x\in\mathcal{X}}}
-\log P\left(  x\right)  U_{1,x}\right)  $ which follows from noting that $%
{\textstyle\sum_{x\in\mathcal{X}}}
U_{1,x}=1$ and can be computed as%
\begin{align}
\text{Var}\left(
{\textstyle\sum_{x\in\mathcal{X}}}
-\log P\left(  x\right)  U_{1,x}\right)    & =\text{Var}\left(
{\textstyle\sum_{x\in\mathcal{X}}}
-\log P\left(  x\right)  1\left(  X_{1}=x\right)   \right)  \\
& =\text{Var}\left(  -\log P\left(  X_1\right)  \right)  \\
& =V(P)
\end{align}
such that
\begin{equation}\label{eq:c2_bound}
c_{2}\leq\frac{400\left\vert \mathcal{X}\right\vert ^3}{V(P)^{3/2}
}\leq\frac{400\left\vert \mathcal{X}\right\vert ^3}{\beta^{3/2}}
\end{equation}
where we have applied the assumption that $V(P)\ge\beta$. The third constant $c_{3}$ is obtained by computing the left side of
(\ref{FinalProb}) using the Gaussian approximation in (\ref{FinalProb}) over
$[H\left(  P\right)  +\sqrt{\frac{V\left(  P\right)  }{n}}\delta-\frac
{1}{\sqrt{n}},\infty)$ and expanding the resulting $Q\left(  \delta-\frac
{1}{\sqrt{nV\left(  P\right)  }}\right)  $ about $\delta$ to obtain%
\begin{equation}\label{eq:c3_bound}
c_{3}=\frac{Q^{\prime}\left(  \delta\right)  }{\sqrt{V\left(  P\right)  }}
\le \frac{1}{\sqrt{2\pi V(P)}}
\end{equation}
where $Q^{\prime}\left(  \delta\right)  $ is the derivative of the $Q$
function evaluated at $\delta$, and the inequality holds because $Q'(\delta)\le\frac{1}{\sqrt{2\pi}}$ for all $\delta$. Combining \eqref{c1_bound_univ}, \eqref{eq:c2_bound}, and \eqref{eq:c3_bound} yields $c_1+c_2+c_3\le B$, where $B$ is given by (\ref{ConstantB}). \ 
\end{IEEEproof}

\subsection{Type Class Size}

Obtaining third-order asymptotic bounds on achievable rates requires precise bounds on the size of type classes. The size of a type class is closely related to the empirical entropy of the type, but importantly one is not strictly increasing with the other. The following Lemma, from an exercise in \cite{CK:book} makes this precise.

\begin{lemma}[Exercise 1.2.2 in \cite{CK:book}]
\label{lemma:type_class_size}The size of the class of type $t$
is bounded as%
\begin{equation}
nf\left(  t\right)  +C^{-}\leq\log|T_{t}|\leq nf\left(  t\right)
\label{typesizebd}%
\end{equation}
where $C^{-}=\frac{1-|\mathcal{X}|}{2}\log(2\pi)-\frac{|\mathcal{X}|}{12\ln 2}$ and
\beq
f(t)= H(t)+\frac{1-\left\vert \mathcal{X}\right\vert }{2n}\log n+\frac{1}%
{2n}\sum\limits_{x\in\mathcal{X}}\min\left\{  \log n,-\log t(x)\right\}  .
\eeq
\end{lemma}

We apply Lemma~\ref{lemma:type_class_size} in combination with Lemma~\ref{lemma:empirical_entropy} to prove the following lemma, giving bounds on the distribution of the size of the type class given by the empirical entropy.

\begin{lemma}\label{lemma:type_size_prb}
Fix $P\in\calP$ such that $V(P)>0$. There exist a finite constant $B$ (dependent on $P$) such that for any $\gamma$
\beq
\left| \bbP\left(\log |T_{t_{X^n}}|>\gamma\right)-Q\left(\frac{\gamma-\frac{1-|\calX_P|}{2}\log n-nH(P)}{\sqrt{nV(P)}}\right)\right|\le\frac{B}{\sqrt{n}}.
\eeq
\end{lemma}

\begin{IEEEproof}
Define the event
\beq
\calE:=\left\{t_{X^n}(x)<\frac{P_X(x)}{2}\text{ for any }x\in\calX_P\right\}.
\eeq
By Chernoff bounds, $\bbP(\calE)\le |\calX_P|e^{-nD}$, where
\beq
D:=\min_{x\in\calX_P}D\left(  P(x)/2\| P(x)\right)>0.
\eeq
We may upper bound the CDF of $\log |T_{t_{X^n}}|$ by
\begin{align}
\bbP\left[\log |T_{t_{X^n}}|>\gamma\right]
&  \leq\mathbb{P}\left[  \log|T_{t(X^{n})}|>\gamma,\calE^c\right]+\bbP[\calE]  \\
&  \leq\mathbb{P}\left[  nf\left(  t_{X^{n}}\right) >\gamma,\calE^c\right] +\bbP[\calE]\label{eq:pstep1}\\
&  \leq\mathbb{P}\left[  nH(t_{X^{n}})+\frac{1-|\calX_P|}{2}\,\log n+\frac{1}{2}\sum_{x\in\calX_P}-\log\frac{P(x)}{2}  >\gamma\right]  +\bbP[\calE]\label{eq:pstep2}
\\&\le Q\left(\frac{\gamma-\frac{1-|\calX_P|}{2}\log n-\frac{1}{2}\sum_{x\in\calX_p}-\log P(x)/2-nH(P)}{\sqrt{nV(P)}}\right)+\frac{B}{\sqrt{n}}+\bbP[\calE]\label{eq:pstep3}
\\&\le Q\left(\frac{\gamma-\frac{1-|\calX_P|}{2}\log n-nH(P)}{\sqrt{nV(P)}}\right)+\frac{1}{2\sqrt{2\pi nV(P)}}\sum_{x\in\calX_P} -\log\frac{P(x)}{2}+\frac{B}{\sqrt{n}}+\bbP[\calE]\label{eq:pstep4}
\end{align}
where (\ref{eq:pstep1}) follows from Lemma~\ref{lemma:type_class_size},  \eqref{eq:pstep2} holds by the definition of $\calE$, \eqref{eq:pstep3} holds by Lemma~\ref{lemma:empirical_entropy}, and \eqref{eq:pstep4} holds because the maximum derivative of $Q$ is $1/\sqrt{2\pi}$. On the other hand, we may lower bound the CDF by
\begin{align}
\bbP\left[\log|T_{t_{X^n}}|>\gamma\right]
&\ge \bbP\left[n f(t_{X^n})+C^->\gamma\right]\label{eq:lstep1}
\\&\ge \bbP\left[ nH(t_{X^n})+\frac{1-|\calX_P|}{2}\log n+C^->\gamma\right]\label{eq:lstep2}
\\&\ge Q\left(\frac{\gamma-\frac{1-|\calX_P|}{2}\log n-C^--nH(P)}{\sqrt{nV(P)}}\right)-\frac{B}{\sqrt{n}}\label{eq:lstep3}
\\&\ge Q\left(\frac{\gamma-\frac{1-|\calX_P|}{2}\log n-nH(P)}{\sqrt{nV(P)}}\right)+\frac{C^-}{\sqrt{2\pi nV(P)}}-\frac{B}{\sqrt{n}}\label{eq:lstep4}
\end{align}
where \eqref{eq:lstep1} holds by Lemma~\ref{lemma:type_class_size}, \eqref{eq:lstep3} holds by Lemma~\ref{lemma:empirical_entropy}, and \eqref{eq:lstep4} holds again by upper bound on the derivative of $Q$. Combining \eqref{eq:pstep4} with \eqref{eq:lstep4} completes the proof.
\end{IEEEproof}

\section{\label{SecAch}Achievable\ Schemes}

\subsection{Two-Stage Codes}

\newcommand{\phis}{\phi_n^{\textsf{2S-FV}}}
\newcommand{\phisf}{\phi_n^{\textsf{2S-FF}}}

A typical approach to encode sequences from an unknown i.i.d. distribution is
to use a two-stage descriptor to encode the type $t$ of the sequence $x^{n}$
first followed by its index within the type class $T_{t}$ \cite[Chap. 13, pp.
433]{Cover:book}. We refer to such a coding scheme as a \emph{Two-Stage code}. There is some variety in the class of Two-Stage codes, depending on the exact choice of first and second stages. We study two specific Two-Stage codes with fixed-length first stages: that is, the number of bits used to express the type of the source sequence is fixed. Let $\phis$ be the $n$-length Two-Stage code with fixed-length first stage and optimal variable-length second stage. That is, given a source sequence with type $t$, the second stage assigns elements of $T_t$ to the shortest $|T_t|$ bit strings in $\{0,1\}^\star$ in any order. It is easy to see that this code  is the optimal two-stage code with fixed-length first stage. Note also that it is not a prefix code. Let $\phisf$ be the $n$-length Two-Stage code with fixed-length first stage and fixed-length second stage, wherein for a source sequence with type $t$, the second stage consists of $\lceil \log |T_t|\rceil$ bits. This code is prefix. The following theorem characterizes the performance of each these two codes.

\newcommand{\kfv}{k_{\textsf{FV}}}
\newcommand{\kff}{k_{\textsf{FF}}}

\begin{theorem}
\label{thm:two-stage} The $\eps$-rates achieved by $\phis$ and $\phisf$ are given by
\begin{align}
R(\phis)&=\frac{1}{n}(s+\kfv(\eps))\label{Rate_Th1}\\
R(\phisf)&=\frac{1}{n}(s+\kff(\eps))\label{Rate_FF}
\end{align}
where
\begin{align}
s&=\left\lceil \log\binom{n+|\mathcal{X}_P|-1}{|\mathcal{X}_P|-1} \right\rceil\\
\kfv(\epsilon)  &=\min\left\{k\in\bbZ:\sum_{t} \mathbb{P}\left(  T_{t}\right)  \left\vert 1-\frac{2^{k+1}-1}{\left\vert T_{t}\right\vert} \right\vert ^{+}\le\epsilon\right\}.\label{eq:k_def}\\
\kff(\epsilon) & =\min\left\{k\in\bbZ:\bbP(\log |T_{t_{X^n}}|> k)\le \eps\right\}.
\end{align}
Moreover, both $R(\phis;\eps,P)$ and $R(\phisf;\eps,P)$ can be written
\begin{equation}\label{eq:two_stage_asymptotic}
H\left(  P\right)  +\sqrt{\frac{V(P)}{n}}Q^{-1}%
(\epsilon)
+\frac{ \left\vert \mathcal{X}_{P}\right\vert -1}{2}
\frac{\log n}{n}+O\left(  \frac{1}{n}\right)  .
\end{equation}
\end{theorem}

\begin{IEEEproof}
The number of types with alphabet $\mathcal{X}_P$ is $\binom{n+|\mathcal{X}_P|-1}{|\mathcal{X}_P|-1}$, thus the number of bits required for the fixed-length first stage in either $\phis$ or $\phisf$ is $s$. In the second stage of $\phis$, using at most $k$ bits one can encode $2^{k+1}-1$ sequences. Thus given that $X^n$ has type $t$, the probability of exceeding $k$ bits in the second stage is
\begin{equation}
\left\vert 1-\frac{2^{k+1}-1}{\left\vert T_{t}\right\vert}\right\vert ^{+}.
\end{equation}
Hence $\kfv(\epsilon)$ as defined in \eqref{eq:k_def} is the minimum number of bits such that the probability of the length of the second stage exceeding $\kfv$ is at most $\epsilon$. This proves \eqref{Rate_Th1}. For $\phisf$, the length of the second stage exceeds $k$ if and only if $\log|T_{X^n}|>k$. Thus $\kff(\eps)$ is the smallest length such that the probability of the second stage exceeding it is at most $\eps$. This proves \eqref{Rate_FF}.

To derive the third-order coding rate, we first note that $s=(|\mathcal{X}_P|-1)\log n+O(1)$. Thus it remains to show that both $\kfv(\eps)$ and $\kff(\eps)$ can be written
\begin{equation}\label{eq:k_asymptotic}
nH+\sqrt{nV}Q^{-1}\left(  \epsilon\right)  +\frac{1-\left\vert
\mathcal{X}_P\right\vert }{2}\log n+O(1).
\end{equation}
This follows for $\kff(\eps)$ directly from Lemma~\ref{lemma:type_size_prb}. Now consider $\kfv(\eps)$, which we may write
\begin{align}
\kfv(\epsilon) 
&  =\min\left\{  k:\mathbb{E}\left\vert 1-(2^{k+1}-1)2^{-Y}\right\vert ^{+}%
\le\epsilon\right\}\label{k_exp}
\end{align}
where $Y:=\log|T_{t_{X^n}}|$. 
Let $g_k(y):=|1-(2^{k+1}-1)2^{-y}|^+$. Since $g_k$ takes values in $[0,1]$ and is monotonically increasing for $y>\log(2^{k+1}-1)$,  the expectation in (\ref{k_exp}) can be written as%
\begin{equation}
\mathbb{E} g_k(Y)=\int_{0}^{1}\mathbb{P}%
\left(  Y>g_k^{-1}\left(  x\right)  \right)  dx.\label{ExpY}%
\end{equation}
We can rewrite the integrand using Lemma~\ref{lemma:type_size_prb} as
\begin{equation}\label{Prh_simplify}
\mathbb{P}\left(  Y>g_k^{-1}\left(  x\right)  \right)  =
\mathbb{P}\left[  g_k\left( nH+\sqrt{nV}Z+\frac{1-\left\vert \mathcal{X}_P\right\vert }%
{2}\log n  \right)  \geq x\right]  +\Theta_{n}(x)%
\end{equation}
where $\left\vert \Theta_{n}(x)\right\vert \leq\frac{B}{\sqrt{n}}$ for all $x$, and
$Z\sim\mathcal{N}\left(  0,1\right)  .$ Let $\Theta_n := \int_0^1 \Theta_n(x) dx$, so $|\Theta_n|\le \frac{B}{\sqrt{n}}$. 
Define
\beq
z_k:=\frac{\log(2^{k+1}-1)-nH-\frac{1-\left\vert \mathcal{X}_P\right\vert }{2}\log n }{\sqrt{nV}}.
\eeq
Now substituting (\ref{Prh_simplify}) in
(\ref{ExpY}) gives
\begin{align}
  \mathbb{E} g_k(Y)
&  =\mathbb{E}\left| 1-(2^{k+1}-1)2^{-nH-\sqrt{nV}Z-\frac{1-\left\vert \mathcal{X}_P%
\right\vert }{2}\log n }\right|  ^{+}  +\Theta_{n}\\
& = \bbE\left| 1-2^{\sqrt{nV}(z_k-Z)}\right|^+ + \Theta_n\\
& = \bbE\left(1-2^{\sqrt{nV}(z_k-Z)}\right)1\left(Z>z_k\right)+\Theta_n\\
& = Q(z_k)- \bbE 2^{\sqrt{nV}(z_k-Z)}1\left(Z>z_k\right)+\Theta_n.\label{eq:EY1}
\end{align}
Let $\Phi_n$ be the second term in \eqref{eq:EY1}. Letting $\varphi$ be the standard Gaussian pdf, for any $\alpha$
\beq
e^{-\alpha z}\varphi(z)=e^{\frac{\alpha^2}{2}} \varphi(z+\alpha).
\eeq
Applying this with $\alpha=(\ln 2)\sqrt{nV}$ gives
\beq
\Phi_n= e^{(\ln 2)\sqrt{nV}z_k+\frac{(\ln 2)^2 nV}{2}} Q\left(z_k+(\ln 2) \sqrt{nV}\right).
\eeq
Using the fact that $Q(x)\le \frac{\varphi(x)}{x}$ for any $x$, we may upper bound $\Phi_n$ by
\begin{align}
\Phi_n
&\le e^{(\ln 2)\sqrt{nV}z_k+\frac{(\ln 2)^2 nV}{2}} \frac{\frac{1}{\sqrt{2\pi}} e^{-(z_k+(\ln 2)\sqrt{nV})^2/2}}{z_k+(\ln 2)\sqrt{nV}}
\\&= \frac{\frac{1}{\sqrt{2\pi}} e^{-z_k^2/2}}{z_k+(\ln 2)\sqrt{nV}}.\label{eq:Phi_bound}
\end{align}
Combining \eqref{eq:Phi_bound} with the fact that $\Phi_n\ge 0$ gives
\beq\label{eq:gk_bounds}
Q(z_k)\le \bbE g_k(Y)\le Q(z_k)+\frac{\frac{1}{\sqrt{2\pi}} e^{-z_k^2/2}}{z_k+(\ln 2)\sqrt{nV}}
\eeq
Recall that $\kfv(\eps)$ is the smallest value of $k$ for which $\bbE g_k(Y)\le\eps$. Define
\begin{align}
k_1&:=\left\lfloor nH+\sqrt{nV}Q^{-1}(\eps)+\frac{1-|\calX_P|}{2}\log n -1\right\rfloor,\\
k_2&:=\left\lceil nH+\sqrt{nV}Q^{-1}(\eps)+\frac{1-|\calX_P|}{2}\log n +d\right\rceil
\end{align}
where $d$ is a constant to be determined. Since $k\le \log(2^{k+1}-1)\le k+1$, we may bound $z_{k_1}\le Q^{-1}(\eps)$, and so by \eqref{eq:gk_bounds}
\beq\label{eq:zk_upper}
\bbE g_{k_1}(Y)\ge Q(z_{k_1})\ge Q\left(\frac{k_1+1-n H-\frac{1-|\calX_P|}{2} \log n}{\sqrt{nV}}\right)\ge 
 \eps.
\eeq
We may also bound
\beq
z_{k_2}\ge Q^{-1}(\eps)+\frac{d}{\sqrt{nV}}.
\eeq
Hence by \eqref{eq:gk_bounds}
\begin{align}
\bbE g_{k_2}(Y)
&\le Q(z_{k_2})+\frac{\frac{1}{\sqrt{2\pi}} e^{-z_{k_2}^2/2}}{z_{k_2}+(\ln 2)\sqrt{nV}}
\\&\le Q\left(Q^{-1}(\eps)+\frac{d}{\sqrt{nV}}\right)+O\left(\frac{1}{\sqrt{n}}\right)
\\&\le \eps\label{eq:zk_lower}
\end{align}
where the last inequality holds for some constant $d$ and sufficiently large $n$. Combining \eqref{eq:zk_upper} and \eqref{eq:zk_lower} we find, for sufficiently large $n$, $k_1\le \kfv(\eps)\le k_2$. This proves that $\kfv(\eps)$ equals \eqref{eq:k_asymptotic}.
\end{IEEEproof}

The third-order coding rate achieved by these Two-Stage codes matches that in our converse for prefix codes in Sec.~\ref{prefix_converse}. Thus $\phisf$ is a near-optimal universal prefix code. Moreover, Theorem~\ref{thm:two-stage} asserts that $\phis$ achieves the same third-order rate as $\phisf$, suggesting that the Two-Stage structure is not suited to optimality in the absence of the prefix constraint. Indeed, the Type Size Code, discussed below, achieves a third order coding rate $\frac{\log n}{n}$ smaller than that of these Two-Stage codes.

\subsection{Type Size Codes}

\begin{figure}
\centerline{\includegraphics[width=.9\textwidth]{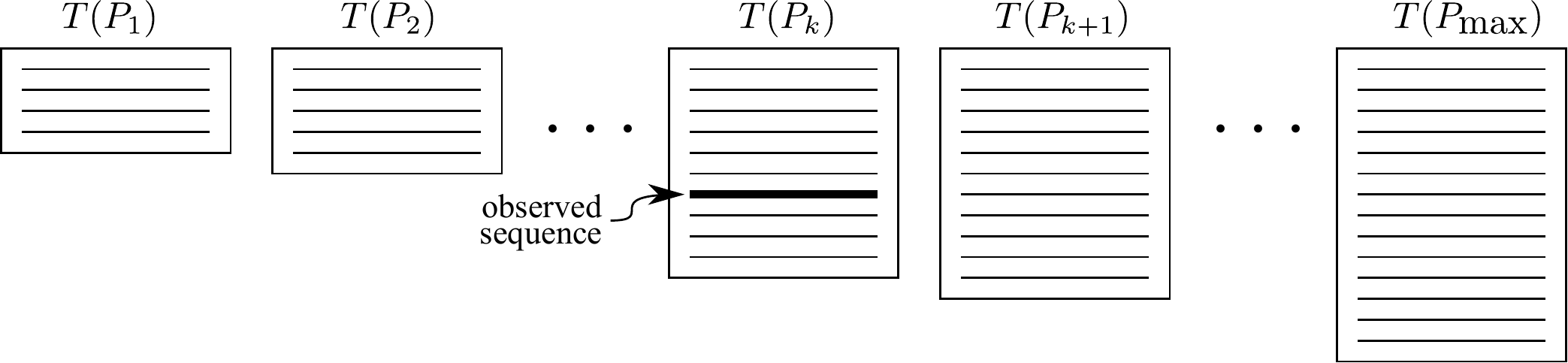}}\caption{Illustration
of the Type Size code. Type classes (denoted $T(P)$ for type $P$) are sorted
from smallest to largest. Given an observed sequence, its codeword is given by
the shortest available bit string, assigned after all previous sequences in
this order.}%
\label{fig:TypeSize}%
\end{figure}

The Type Size Code is illustrated in Fig.~\ref{fig:TypeSize}. Recall that $\calX_{t_{X^{n}}}$ is the support set under $t_{X^{n}}$. The encoding function $\phi$ outputs two strings:

\begin{enumerate}
\item a string of $|\mathcal{X}|$ bits recording $\calX_{t_{X^{n}}}$,
i.e. the elements of $\mathcal{X}$ that appear in the observed sequence, and

\item a string that assigns sequences to $\{0,1\}^{\star}$ in order based on
the size of the type class of the type of $X^{n}$, among all types $t$ with
$\calX_t=\calX_{t_{X^{n}}}$. That is, if $\calX_{t_{x^{n}%
}}=\calX_{t_{x^{\prime n}}}$ and $|T_{t_{x^{n}}}|<|T_{t_{x^{\prime n}}}%
|$, then $\ell(\phi(x^{n}))\leq\ell(\phi(x^{\prime n}))$.
\end{enumerate}

Note that the code described in Section~\ref{SecBinary} for binary sources is
very similar to the Type Size Code. The support set string is omitted, and
type classes with the same size are interleaved rather than ordered one after
the other, but in essential aspects the codes are the same.

\newcommand{\phit}{\phi_n^{\textsf{TS}}}

\begin{theorem}\label{TypeSizeCode}
Let $\phit$ be the $n$-length Type Size Code. It achieves the rate function
\beq\label{eq:type_size_rate}
R(\phit;\eps,P)=\frac{|\calX|}{n}+\frac{1}{n}\left\lfloor\log \min\left\{M:\sum_{\bar\calX\subseteq\calX} \bbP(\calX_{t_{X^n}}=\bar\calX)\eps(\bar\calX,M)\le\eps\right\}\right\rfloor
\eeq
where
\beq\label{eq:eps_M}
\eps(\bar\calX,M)=1-\bbP\big(|T_{t_{X^n}}|<\tau^\star(\bar\calX)|\calX_{t_{X^n}}=\bar\calX\big)-\lambda^\star(\bar\calX)\bbP\big(|T_{t_{X^n}}|=\tau^\star(\bar\calX)|\calX_{t_{X^n}}=\bar\calX\big)
\eeq
where $\tau^\star(\bar\calX)\in\bbN$ and $\lambda^\star(\bar\calX)\in [0,1)$ are chosen so that
\beq\label{eq:eps_M2}
\sum_{\substack{t:|T_t|<\tau^\star(\bar\calX),\\ \calX_t=\bar\calX}} |T_t|
+\lambda^\star(\bar\calX)\sum_{\substack{t:|T_t|=\tau^\star(\bar\calX),\\\calX_t=\bar\calX}}|T_t|=M.
\eeq
\end{theorem}

\begin{IEEEproof}
In the construction of the Type Size Code, for each $\bar\calX\subseteq\calX$, all sequences $x^n$ with $\calX_{t_{x^n}}=\bar\calX$ are sorted by type class size. For each $x^n$, let $m(x^n)$ be the position of $x^n$ in this order. Thus $\ell(\phit(x^n))=|\calX|+\lfloor \log m(x^n)\rfloor$. From the definition of $\eps$-rate, we have
\begin{align}
R(\phit;\eps,P)
&=\min\left\{\frac{k}{n}:\bbP(\ell(\phit(X^n))>k)\le\eps\right\}
\\&=\min\left\{\frac{k}{n}:\bbP(|\calX|+\lfloor\log m(X^n)\rfloor>k)\le\eps\right\}
\\&=\min\left\{\frac{|\calX|+\lfloor\log M\rfloor}{n}:\bbP(\lfloor\log m(X^n)\rfloor>\lfloor\log M\rfloor)\le\eps\right\}
\\&=\frac{|\calX|}{n}+\frac{1}{n}\left\lfloor\log \min\left\{M:\bbP(m(X^n)>M)\le\eps\right\}\right\rfloor.
\end{align}
Moreover
\begin{align}
\bbP(m(X^n)>M)
&=\sum_{\bar\calX\subseteq\calX} \bbP(\calX_{t_{X^n}}=\bar\calX) \bbP(m(X^n)>M|\calX_{t_{X^n}}=\bar\calX)
\\&=\sum_{\bar\calX\subseteq\calX} \bbP(\calX_{t_{X^n}}=\bar\calX) \eps(\bar\calX,M).
\end{align}
This completes the proof.
\end{IEEEproof}

\begin{remark}
Using the standard equivalence between codes and guessing functions described in Sec.~\ref{Guessing}, one can construct a guessing function equivalent to the Type Size Code that achieves 
\beq
M(\eps,P)=2^{|\calX|} \min\left\{M:\sum_{\bar\calX\subseteq\calX} \bbP(\calX_{t_{X^n}}=\bar\calX)\eps(\bar\calX,M)\le\eps\right\}
\eeq
where $\eps(\bar\calX,M)$ is as defined in Theorem~\ref{TypeSizeCode}.
\end{remark}

The following theorem bounds the asymptotic rate achieved by the Type Size Code.

\begin{theorem}
The rate function achieved by the Type Size Code satisfies
\begin{equation}
R(\phit;\epsilon,P) \le H(P)+\sqrt{\frac{V(P)}{n}}Q^{-1}(\epsilon)
+\frac{\left\vert \calX_P\right\vert -3}{2}
\frac{\log n}{n}+O\left(  \frac{1}{n}\right).  \label{Rapprox}%
\end{equation}
\end{theorem}

\begin{IEEEproof}
Fix $\eps$ and $P$. Let $M^\star$ be the minimizing $M$ in the second term in \eqref{eq:type_size_rate}. Thus, Theorem~\ref{TypeSizeCode} can be written
\beq\label{eq:R_ts_bd1}
R(\phit;\eps,P)=\frac{|\calX|}{n}+\frac{1}{n}\lfloor \log M^\star\rfloor.
\eeq
We proceed to upper bound $M^\star$, beginning by  upper bounding the sum over $\bar\calX$ inside the second term in \eqref{eq:type_size_rate}. Let $\mu_n=\bbP(\calX_{t_{X^n}}\ne \calX_P)$.  Large deviation bounds can be used to derive that  $\mu_n$ vanishes exponentially fast in $n$. Thus
\begin{align}
\sum_{\bar\calX\subseteq\calX} \bbP(\calX_{t_{X^n}}=\bar\calX)\eps(\bar\calX,M^\star)
&\le \eps(\calX_P,M^\star)+\mu_n
\\&\le \bbP(|T_{t_{X^n}}|\ge \tau^\star(\calX_P))+\mu_n.
\end{align}
Since $\eps(\calX_P,M^\star)$ is decreasing in $\tau^\star(\calX_P)$, for any real number $\tau$ satisfying $\bbP(|T_{t_{X^n}}|\ge \tau)\le \eps-\mu_n$, it must be that $\tau\ge \tau^\star(\calX_P)$. Thus, applying the definition of $\tau^\star(\calX_P)$, we have
\begin{align}
M^\star
&\le \sum_{\substack{t:|T_t|\le \tau^\star(\calX_P)\\ \calX_t=\calX_P}} |T_t|
\\&\le \min_{\tau:\bbP(|T_{t_{X^n}}|\ge \tau)\le \eps-\mu_n} \sum_{\substack{t:|T_t|\le \tau \\ \calX_t=\calX_P}} |T_t|.
\end{align}

By Lemma~\ref{lemma:type_size_prb} there exists a constant $B$ such that
\beq
\bbP( |T_{t_{X^n}}|\ge \tau)
\le Q\left(  \sqrt{\frac{n}{V(P)}}\left[ \frac{\log\tau}{n}-\frac{1-|\calX_P|}{2}\,\frac{\log n}{n}-H(P)\right]  \right)
+\frac{B}{\sqrt{n}}.
\eeq
Hence, if we define $\tau^\star$ such that\footnote{Note that $\tau^\star$ is not quite the same as $\tau^\star(\calX_P)$.}
\beq\label{eq:tau_star}
\frac{\log\tau^\star}{n}=H(P)+\sqrt{\frac{V(P)}{n}}Q^{-1}\left(\eps-\mu_n-\frac{B}{\sqrt{n}}\right)+\frac{1-|\calX_P|}{2}\frac{\log n}{n}.
\eeq
then $\bbP(|T_{t_{X^n}}|\ge\tau^\star)\le \eps-\mu_n$. Thus 
\beq
M^\star\le \sum_{\substack{t:|T_t|\le\tau^\star\\ \calX_t=\calX_P}} |T_t|.
\eeq

 Let $\gamma^\star:=\frac{\log \tau^\star}{n}$. Also fix $\Delta>0$, and, for integers $i$, define $a_i=\gamma^\star-C^-/n-i\Delta$ and $\calA_i=\{P\in\calP:a_i-\Delta<f(P)\le a_i\}$. We may write
\begin{align}
M^\star
&\le\sum_{\substack{t:\frac{1}{n}\log |T_t|\le\gamma^\star\\ \calX_t=\calX_P}} |T_t|
\\&\leq\sum_{\substack{t:f(t)+\frac
{C^{-}}{n}\leq\gamma^\star\\\calX_t=\calX_P}}2^{nf(t)}\label{eq:M_eps1}\\
& =\sum_{i=0}^{\infty}\sum_{\substack{t\in\calA_i\cap\calP_n\\\calX_t=\calX_P}}2^{nf(t)}\\
& \leq\sum_{i=0}^{\infty}|\calA_i\cap\calP_n\cap\calP_{\calX_P}| 2^{na_i}.\label{eq:mbdstart}%
\end{align}
where in \eqref{eq:M_eps1} we have applied Lemma~\ref{lemma:type_class_size} in two different ways, and in \eqref{eq:mbdstart} we have defined $\calP_{\calX_P}$ as the set of distributions with support set $\calX_P$. We now bound the term $|\calA_i\cap\calP_n\cap\calP_{\calX_P}|$. Define a 2-norm ball of radius $1/2n$ around a distribution $P$ as
$B(P)=\left\{  Q:\Vert P-Q\Vert_{2}<\frac{1}{2n}\right\}  $. Note that for any
two different types $t_{1},t_{2}$, $\Vert t_{1}-t_{2}\Vert_{2}\geq\frac{1}{n}$,
so $B(t_{1})$ and $B(t_{2})$ are always disjoint. Since $\mathcal{P}_{\calX_P}$ is an
$(|\calX_P|-1)$-dimensional space, we define volumes on $\mathcal{P}%
_{\calX_P}$ via the $(|\calX_P|-1)$-dimensional Lebesgue
measure. For any type $t\in\mathcal{P}_{\calX_P}$, $t(x)\geq1/n$ for
all $x\in\calX_P$, so
\begin{equation}
\text{Vol}(B(t)\cap\mathcal{P}_{\calX_P})=d/n^{|\calX_P|-1}%
\end{equation}
for a constant $d$ that depends only on $|\calX_P|$. We may bound the number of types in $\calA_i$ with $\calX_t=\calX_P$ by
\begin{align}
|\calA_i\cap\mathcal{P}_{n}\cap\mathcal{P}_{\calX_P}| 
&=\sum_{t\in\calA_i\cap\mathcal{P}_{n}\cap\mathcal{P}_{\calX_P}%
}\frac{n^{|{\calX_P}|-1}}{d}\text{Vol}(B(t)\cap\mathcal{P}%
_{\calX_P})\\
&  =\frac{n^{|\calX_P|-1}}{d}\text{Vol}\left(
{\bigcup_{t\in\calA_i\cap\mathcal{P}_{n}\cap\mathcal{P}_{\calX_P}}}
B(t)\cap\mathcal{P}_{\calX_P}\right)  \label{eq:voldisjoint}\\
& \leq\frac{n^{|\calX_P|-1}}{d}\text{Vol}\left(
{\bigcup_{Q\in\calA_i}}
B(Q)\cap\mathcal{P}_{\calX_P}\right)  \label{eq:volbd}%
\end{align}
where (\ref{eq:voldisjoint}) holds because the balls are disjoint. There
exists a constant $C$ so that for any distributions $Q_1$ and $Q_2$,
\begin{equation}
|f(Q_1)-f(Q_2)|\leq C\Vert Q_1-Q_2\Vert_{2}.
\end{equation}
In particular, for any $Q_1\in B(Q_2)$, 
\begin{equation}|f(Q_1)-f(Q_2)|\leq C/2n.\label{fdist}%
\end{equation}
For any
$\lambda\geq0$ let
\begin{equation}
g(\lambda)=\text{Vol}(\{Q\in\mathcal{P}_{\calX_P}:f(Q)\leq
\lambda\}).\label{hdef}%
\end{equation}
Let $K$ be the constant so that for all $a,b$,
\begin{equation}
|g(a)-g(b)|\leq K|a-b|.\label{Kdef}%
\end{equation}
Note that $K$ depends only on $|\calX_P|$. For any real
$a$,
\begin{align}
|\calA_i\cap \calP_n\cap \calP_{\calX_P}|
&  \leq\frac{n^{|\calX_P|-1}}{d}\text{Vol}\left(  \bigcup
_{Q:a_i-\Delta<f(Q)\leq a_i}B(Q)\cap\mathcal{P}_{\calX_P}\right)
\label{eq:ntbd1}\\
&  \leq\frac{n^{|\calX_P|-1}}{d}\text{Vol}\left(  \left\{
Q\in\mathcal{P}_{\calX_P}:f(Q)\textstyle\in(a_i-\frac{C}{2n}-\Delta,a_i+\frac{C}{2n}]\right\}
\right)  \label{eq:ntbd2}\\
&  =\frac{n^{|\calX_P|-1}}{d}\left[  g\left(  a_i+\frac{C}%
{2n}\right)  -g\left(  a_i-\frac{C}{2n}-\Delta\right)  \right]  \label{eq:ntbd3}\\
&  \leq\frac{Kn^{|\calX_P|-1}}{d}\left[  \Delta+\frac{C}{n}\right]
\label{eq:ntbd4}%
\end{align}
where (\ref{eq:ntbd1}) holds by (\ref{eq:volbd}), (\ref{eq:ntbd2}) holds by
(\ref{fdist}), (\ref{eq:ntbd3}) holds by the definition of $h$ in
(\ref{hdef}), and (\ref{eq:ntbd4}) holds
by (\ref{Kdef}).

Applying  (\ref{eq:ntbd4})  to (\ref{eq:mbdstart}), we
obtain%
\begin{align*}
M^\star
 & \leq\sum_{i=0}^{\infty}\frac{Kn^{|\calX_P|-1}}%
{d}\left[  \Delta+\frac{C}{n}\right]  \exp\{n\gamma^\star-C^{-}-ni\Delta\}\\
& =\frac{Kn^{|\calX_P|-1}}{d}\left[  \Delta+\frac{C}{n}\right]
\frac{\exp\{n\gamma^\star-C^{-}\}}{1-\exp\{-n\Delta\}}.
\end{align*}
The above holds for any $\Delta>0$, so we may take $\Delta=\frac{C}{n}$ to
write
\begin{align}
\log M^\star &  \leq n\gamma^\star-C^{-}
+\log\left[  \frac{Kn^{|\calX_P|-1}}{d}\frac{2C}
{n}\frac{1}{1-\exp\{-C\}}\right]  \\
&  =n\gamma^\star+(|\calX_P|-2)\log n-C^{-}+\log\left[  \frac{2KC}{d(1-\exp\{-C\})}\right] \\
&  =nH(P_{X})+\sqrt{nV(P)}Q^{-1}(\epsilon)+\frac{|\calX_P|-3}{2}\log n+O(1)\label{eq:M_eps_final}
\end{align}
where we have used the expression for $\tau^\star$ (and equivalently $\gamma^\star$) from \eqref{eq:tau_star}, as well as the fact $Q^{-1}(\eps-\mu_n-\frac{B}{\sqrt{n}})=Q^{-1}(\eps)+O(\frac{1}{\sqrt{n}})$, since $\mu_n$ is exponentially decreasing. Applying \eqref{eq:M_eps_final} to \eqref{eq:R_ts_bd1} completes the proof.
\end{IEEEproof}

\section{\label{SecConv}Converse Results}

In this section, we develop tight outer bounds on the third order coding rate
of fixed-to-variable length coding schemes for both general codes and prefix codes. Intuitively, our converse bounds
arise from the degree of uncertainty about the source distribution. What the
bound reveals is that if the set of distributions that occur has dimension $d$
$(=\left\vert \mathcal{X}_{P}\right\vert -1)$, then the required rate for the
universal code with be approximately $\frac{d}{2}\frac{\log n}{n}$ larger than
in the non-universal setting.

Consider a specific source distribution $P_{0}$. In the non-prefix setting,
what matters is uncertainty among $P_{0}$ and other distributions with
approximately the same entropy. This is because the `natural' length of
codewords for typical sequences drawn from distribution $P$ is about $nH(P)$.
Thus sequences with $H(P)\approx H(P_{0})$ compete with each other for the
same codewords. Distributions with substantially different entropy have little
effect each other. The dimension of this set is $d=|\mathcal{X}|-2$. This
dimension leads to a converse bound on the rate of about $\frac{|\mathcal{X}%
|-2}{2}\frac{\log n}{n}$ larger than in the non-universal setting (i.e., a third-order coefficient of $\frac{|\calX|-3}{2}$). This is
precisely the third-order coding rate achieved by the Type Size code,
indicating that the Type Size code performs about as well as any universal scheme.

Our converse proof for general codes makes use of the bounds on the distribution of the empirical entropy derived in Lemma~\ref{lemma:empirical_entropy}, as well as an application of Laplace's approximation, as described next in Sec.~\ref{laplace}. In Sec.~\ref{mixture}, we apply Laplace's approximation to bound the values of mixture distributions, which will be a key element in our converse proofs. Our converse for general fixed-to-variable codes is presented in Sec.~\ref{general_converse}, and for prefix codes in Sec.~\ref{prefix_converse}.

\subsection{Laplace's Approximation}\label{laplace}

Laplace's approximation allows one to approximate an integral around the
maximum of the integrand on both vector spaces and manifolds. The following theorem gives the result for integrals on $\bbR^k$. Subsequently, Corollary~\ref{corollary:laplace_manifold} extends the result to integrals on manifolds.
\begin{theorem}[\cite{Wong:Book89}, Chap. 9, Thm. 3]\label{thm:laplace}
Let $D\subset \bbR^k$, and $f$ and $g$ be functions that are infinitely differentiable on $D$. Let
\beq
J( n)=\int_D g(x)e^{- n f(x)}dx.
\eeq
Assume that
\begin{enumerate}
\item The integral $J( n)$ converges absolutely for all $ n\ge  n_0$.
\item There exists a point $x^\star$ in the interior of $D$ such that for every $\eps>0$, $\rho(\eps)>0$ where
\beq
\rho(\eps)=\inf\{f(x)-f(x^\star):x\in D\text{ and }|x-x^\star|\ge \eps\}.
\eeq
\item The Hessian matrix
\beq
A=\left(\frac{\partial^2 f}{\partial x_i\partial x_j}\right)\Big|_{x=x^\star}
\eeq
is positive definite.
\end{enumerate}
Then
\beq
J( n)=e^{- n f(x^\star)}\left(\frac{2\pi}{ n}\right)^{k/2} g(x^\star)|A|^{-1/2}\left(1+O( n^{-1})\right).
\eeq
\end{theorem}
\begin{corollary}\label{corollary:laplace_manifold}
Let $D$ be a $k$-dimensional differentiable manifold embedded in $\bbR^{m}$. Consider the same setup as Theorem~\ref{thm:laplace}. Let $F\in\bbR^{m\times k}$ be an orthonormal basis for the tangent space to $D$ at $x^\star$. Then
\[
J( n)=e^{- n f(x^\star)}\left(\frac{2\pi}{ n}\right)^{k/2} g(x^\star)|F^TAF|^{-1/2} \left(1+O( n^{-1})\right).
\]
\end{corollary}
\begin{IEEEproof}
Define a function $h:\bbR^{k}\to D$ as
\beq
h(y):=\argmin_{x\in D}\|x-(x^\star+Fy)\|_2.
\eeq
Since $D$ is a differentiable manifold, there exists a neighborhood $U\subset D$ of $x^\star$ on which $h$ is a diffeomorphism. Moreover, $h'(0)=F$. By changing variables using $h$ and applying Theorem~\ref{thm:laplace}, we find
\begin{align}
\int_U g(x)e^{- n f(x)}dx &= \int_{h^{-1}(U)} g(h(y)) |h'(y)^Th'(y)| e^{- n f(h(y))} dy\\
&=e^{- n f(x^\star)}\left(\frac{2\pi}{ n}\right)^{k/2}g(x^\star)|h'(0)^Th'(0)|\,|h'(0)^T A h'(0)|^{-1/2}\left(1+O( n^{-1})\right)\\
&=e^{- n f(x^\star)}\left(\frac{2\pi}{ n}\right)^{k/2}g(x^\star)|F^TAF|^{-1/2} \left(1+O( n^{-1})\right)\label{eq:int_U}
\end{align}
where we have used the fact that $F^TF=I$ because the columns of $F$ are orthornormal. It is easy to see that there exist constants $K$ and $\delta>0$ such that
\beq\label{eq:int_not_U}
\int_{D\setminus U}g(x)e^{- n f(x)}dx \le Ke^{- n(f(x^\star)+\delta)}.
\eeq
Combining \eqref{eq:int_U} with \eqref{eq:int_not_U} completes the proof.
\end{IEEEproof}

\subsection{Approximating Mixture Distributions}\label{mixture}

The following lemma on mixture distributions uses Theorem~\ref{corollary:laplace_manifold} and bounds the distribution of a uniform
mixture of i.i.d. distributions.

\begin{lemma}
\label{lemma:mixture_bound} Let $\mathcal{P}_{0}$ be a subset of the
probability simplex on $\mathcal{X}$ that is a $k$-dimensional
differentiable manifold, and let $\bar{P}(x^{n})$ be a uniform mixture among 
$n$-length i.i.d.~distributions with marginals in $\mathcal{P}_{0}$. That is 
\begin{equation}
\bar{P}(x^{n})=\frac{1}{\text{Vol}(\mathcal{P}_{0})}\int_{P\in \mathcal{P}%
_{0}}P^{n}(x^{n})dP.
\end{equation}%
Let $\bar\calX=\bigcup_{P\in\calP_0} \calX_P$. For any sequence $x^{n}$, let $p_{\min }(x^{n}):=\min_{x\in 
\bar\calX}t_{x^{n}}(x)$. Assume that there is a unique 
\begin{equation}
P^{\star }=\argmin_{P\in \mathcal{P}_{0}}D(t_{x^{n}}\Vert P).
\end{equation}%
Then 
\begin{equation}
\bar{P}(x^{n})\leq \frac{2^{-nH(t_{x^{n}})}}{\text{Vol}(\mathcal{P}_{0})}%
\left( \frac{2\pi }{p_{\min }(x^{n})n}\right) ^{k/2}\big(1+O(n^{-1})\big).
\label{eq:mixture_bound}
\end{equation}
\end{lemma}

\begin{IEEEproof}
Fix a sequence $x^{n}$ with type $t$. If $t(x)>0$ for any $x\notin \bar\calX$, then certainly $\bar{P}(x^{n})=0$, so \eqref{eq:mixture_bound}
holds. We henceforth assume that $t(x)=0$ for all $x\notin \bar\calX$.
We have 
\begin{equation}
\bar{P}(x^{n})=\frac{1}{\text{Vol}(\mathcal{P}_{0})}\int_{P\in \mathcal{P}%
_{0}}2^{-n(H(t)+D(t\Vert P))}dP.  \label{eq:P_bar_int}
\end{equation}%
If $P^{\star }$ is on the boundary of $\mathcal{P}_{0}$, extend $\mathcal{P}%
_{0}$ so that it is remains a $k$-dimensional manifold but with $P^{\star }$
in its interior, where $P^{\star }$ is still the unique minimizer of $%
D(t\Vert P)$ for $P\in \mathcal{P}_{0}$. Thus 
\begin{equation}
\bar{P}(x^{n})\leq \frac{1}{\text{Vol}(\mathcal{P}_{0})}\int_{\mathcal{P}%
_{0}}2^{-n(H(t)+D(t\Vert P))}dP.  \label{eq:P_bar_int2}
\end{equation}%
Applying Corollary~\ref{corollary:laplace_manifold} to the integral in %
\eqref{eq:P_bar_int2} with $f(P)=(\ln 2)D(t\Vert P)$ and $g(P)=1$ gives 
\begin{equation}
\bar{P}(x^{n})\leq \frac{2^{-n(H(t)+D(t\Vert P^{\star }))}}{\text{Vol}(%
\mathcal{P}_{0})}\left( \frac{2\pi }{n}\right) ^{k/2}|F^{T}AF|^{-1/2}\left(
1+O(n^{-1})\right)   \label{eq:mixture_bound1}
\end{equation}%
where $F$ is an orthonormal basis for the tangent space to $\mathcal{P}_{0}$
at $P^{\star }$, and $A$ is an $|\bar\calX|\times |\bar\calX|$
diagonal matrix with elements $\frac{t(x)}{P^{\star }(x)^{2}}$. We lower
bound the singular values of $F^{T}AF$ as follows. Take any $\mathbf{y}$
with $\Vert \mathbf{y}\Vert =1$, and we have (letting $\mathbf{z}=F\mathbf{y}
$) 
\begin{align}
\sigma _{i}(F^{T}AF)& \geq \Vert F^{T}AF\mathbf{y}\Vert  \\
& \geq \mathbf{y}^{T}F^{T}AF\mathbf{y} \\
& =\mathbf{z}^{T}A\mathbf{z} \\
& \geq \Vert \mathbf{z}\Vert \min_{x\in \bar\calX}\frac{t(x)}{P^{\star
}(x)^{2}} \\
& =\min_{x\in \bar\calX}\frac{t(x)}{P^{\star }(x)^{2}}
\label{eq:sigma_bd_1} \\
& \geq p_{\min }(x^{n})  \label{eq:sigma_bd_2}
\end{align}%
where in \eqref{eq:sigma_bd_1} we have used the fact that $\Vert F\mathbf{y}%
\Vert =\Vert \mathbf{y}\Vert =1$ by the orthonormality of the columns of $F$%
, and in \eqref{eq:sigma_bd_2} we have used that $P^{\star }(x)\leq 1$ and
the definition of $p_{\min }(x^{n})$. Now we have that 
\begin{equation}
|F^{T}AF|=\prod_{i=1}^{k}\sigma _{i}(F^{T}A^{-1}F)\geq p_{\min }(x^{n})^{k}
\end{equation}%
Applying this to \eqref{eq:mixture_bound1} and using the fact that $D(t\Vert
P^{\star })\geq 0$ proves \eqref{eq:mixture_bound}.
\end{IEEEproof}

\begin{remark}
The crux of the statement of Lemma~\ref{lemma:mixture_bound} is the exponent $k/2$. In applying this lemma, $k=|\mathcal{X}|-2$ is the dimension of uncertainty in the probability distributions, so this yields a bound on the third-order coding rate $\frac{k}{2} \frac{\log n}{n}$ larger than that of non-universal codes.
\end{remark}

\subsection{Converse Bound for General Fixed-to-Variable Codes}\label{general_converse}

The following is a simple finite blocklength converse bound.

\begin{theorem}
\label{thm:converse} Fix any set $\mathcal{P}_{0}$ of distributions on
$\mathcal{X}$, and let $\bar{P}(x^{n})$ be any mixture distribution of
$n$-length i.i.d.~distributions with marginals in $\mathcal{P}_{0}$. For any $n$-length code $\phi$, if we set
\begin{equation}
k := \max_{P\in\mathcal{P}_{0}} nR(\phi;\epsilon,P)
\end{equation}
then
\begin{equation}
\epsilon\ge\max_{\tau>0} \bar{\mathbb{P}}(-\log\bar{P}(X^{n})\ge
k+\tau)-2^{-\tau}.
\end{equation}

\end{theorem}

\begin{IEEEproof}
By definition of $R(n,\epsilon,P)$, we have
\begin{equation}
\mathbb{P}(\ell(\phi(X^{n}))\geq k)\leq\epsilon,\qquad\text{for all }%
P\in\mathcal{P}_{0}.
\end{equation}
Certainly
$
\bar{\mathbb{P}}(\ell(\phi(X^{n}))\geq k)\leq\epsilon.
$
Using Theorem~3 in \cite{YannisVerdu}, for any $\tau>0$, we obtain%
\begin{equation}
\bar{\mathbb{P}}(\ell(\phi(X^{n}))\geq k)\geq\bar{\mathbb{P}}(-\log\bar{P}%
(X^{n})\geq k+\tau)-2^{-\tau}.
\end{equation}
\end{IEEEproof}

Now we use Theorem~\ref{thm:converse} to derive the following converse bound
on the third-order coding rate. Define $J_{\eps,n}(P):=H(P)+\sqrt{\frac{V(P)}{n}} Q^{-1}(\eps)$. When the relevant values of $\eps$ and $n$ are clear from context, we write simply $J(P)$.

\begin{theorem}\label{thm:general_converse}
Fix $\bar\calX\subset\calX$, $\eps>0$, and $\Gamma\in(0,\log|\bar\calX|)$. There exists a finite constant $d_{\Gamma}$ such that, for any blocklength $n$ and any $n$-length code $\phi_n$,
\beq
\sup_{\substack{P:\calX_P=\bar\calX,\\ J(P)=\Gamma}} R(\phi_n;\eps,P)\ge \Gamma+\frac{|\bar\calX|-3}{2}\frac{\log n}{n}-\frac{d_\Gamma}{n}.
\eeq 
\end{theorem}

Before proving the theorem, we provide the following straightforward corollary.

\begin{corollary}
\label{thm:converse_third_order} For any $\bar\calX\subset\mathcal{X}$, $\epsilon>0$, and any sequence of codes $\phi_n$,
\begin{equation}
\sup_{P:\calX_P=\bar\calX}\left[  R(\phi_n;\epsilon
,P)-H(P)-\sqrt{\frac{V(P)}{n}}Q^{-1}(\epsilon)\right] 
\geq\frac{|\bar\calX|-3}{2}\frac{\log n}{n}-O\left(  \frac{1}{n}\right)
.
\end{equation}
\end{corollary}

\begin{IEEEproof}[Proof of Theorem~\ref{thm:general_converse}]
Let $P_1$ be a constant distribution on an element of $\bar\calX$, and let $P_2$ be a uniform distribution on $\bar\calX$. Note that $J(P_1)=0$ and $J(P_2)=\log|\bar\calX|$. Moreover, since $J$ is a continuous function of $P$, by the intermediate value theorem any continuous path of distributions between $P_1$ and $P_2$ passes through all values of $J$ between $0$ and $\log|\bar\calX|$. Hence, since $0<\Gamma<\log|\bar\calX|$, the set $\{P:\calX_P=\bar\calX,\ J(P)=\Gamma\}$ is a $|\bar\calX|-2$-dimensional manifold. We further choose $\beta>0$ small enough so that
\beq
\calP_0:=\{P:\calX_P=\bar\calX,\ J(P)=\Gamma,\ P(x)\ge\beta\text{ for all }x\in\bar\calX\}
\eeq
is also a $|\bar\calX|-2$-dimensional manifold. Let
\beq
k=\sum_{P\in\calP_0} nR(\phi_n;\eps,p).
\eeq
It suffices to show that there exists finite $d_\Gamma$ such that for any $n$,
\begin{equation}\label{eq:to_prove}
k\geq n\Gamma+
\frac{|\bar\calX|-3}{2}\log n-d_\Gamma.
\end{equation}
Applying Theorem~\ref{thm:converse} gives that, if $\barP$ is a uniform mixture among $n$-length i.i.d. distributions with marginals in $\calP_0$, then
\begin{align}
\epsilon &  \geq\frac{1}{\text{Vol}(\mathcal{P}_{0})}\int_{\mathcal{P}_{0}%
}\mathbb{P}(-\log\bar{P}(X^{n})\geq k+\tau)dP-2^{-\tau}\\
&  \geq\inf_{P\in\mathcal{P}_{0}}\mathbb{P}(-\log\bar{P}(X^{n})\geq
k+\tau)-2^{-\tau}. \label{eq:eps_bd1}%
\end{align}
Let $\mathcal{P}_{0,\delta}:=\{t:\min_{P\in\mathcal{P}_{0}}D(t\Vert
P)\leq\delta\}$. Because $\mathcal{P}_{0}$ has limited curvature, for
sufficiently small $\delta$, if $t(x^{n})\in\mathcal{P}_{0,\delta}$, then
there is a unique $P^{\star}\in\mathcal{P}_{0,\delta}$ minimizing $D(t_{x^{n}%
}||P)$. Also for sufficiently small $\delta$, if $t(x^{n})\in\mathcal{P}%
_{0,\delta}$, then $\min_{x\in\bar\calX}t_{X^{n}}(x)\geq\beta/2$.
Choose $\delta>0$ small enough to satisfy these two conditions. By
Lemma~\ref{lemma:mixture_bound}, if $t_{x^{n}}\in\mathcal{P}_{0,\delta}$, then
for sufficiently large $n$ (recall $P_{0}$ is ($|\bar\calX|-2$%
)-dimensional)
\begin{equation}
-\log\bar{P}(x^{n})\geq nH(t_{x^{n}})+\frac{|\bar\calX|-2}{2}\log n+c
\label{P_bar_bound}%
\end{equation}
where
\begin{equation}
c=2\log\left\{  \frac{1}{\text{Vol}(\mathcal{P}_{0})}\left(  \frac{4\pi
}{\beta}\right)  ^{\frac{|\bar\calX|-2}{2}}\right\}  .
\end{equation}
Moreover, by Sanov's theorem, for sufficiently large $n$, for any
$P\in\mathcal{P}_{0}$
\begin{equation}
\mathbb{P}(t_{X^{n}}\notin\mathcal{P}_{0,\delta})\leq2^{-n\delta/2}.
\end{equation}
Thus, continuing from \eqref{eq:eps_bd1} and applying (\ref{P_bar_bound}) gives
\begin{align}
\eps+2^{-\tau}&  \geq\inf_{P\in\mathcal{P}_{0}}\mathbb{P}\left(  -\log\bar{P}(X^{n})\geq
k+\tau,t_{X^{n}}\in\mathcal{P}_{0,\delta}\right) \\
&  \geq\inf_{P\in\mathcal{P}_{0}}
\mathbb{P}\left(  nH(t_{X^{n}})+\frac{|\bar\calX|-2}{2}\log n+c\geq
k+\tau,t_{X^{n}}\in\mathcal{P}_{0,\delta}\right) \label{eq:P_star_bd1}\\
&  \geq\inf_{P\in\mathcal{P}_{0}}\mathbb{P}\left(  nH(t_{X^{n}})+\frac
{|\bar\calX|-2}{2}\log n+c\geq k+\tau\right) -2^{-n\delta/2}\\
&  \geq\inf_{P\in\mathcal{P}_{0}}Q\bigg(  \sqrt{\frac{n}{V(P)}}\bigg[
\frac{k+\tau-c}{n}-\frac{|\bar\calX|-2}{2}\frac{\log n}{n}-H(P)\bigg]
\bigg) -\frac{B}{\sqrt{n}}-2^{-n\delta/2}\label{eq:P_star_bd2}
\end{align}
where in \eqref{eq:P_star_bd2} we have applied
Lemma~\ref{lemma:empirical_entropy}. Setting $\tau=\frac{1}{2}\log n$ and  rearranging gives
\begin{align}
k &  \geq\inf_{P\in\mathcal{P}_{0}}nH(P)+\sqrt{n V(P)}%
Q^{-1}\left(  \epsilon+\frac{B+1}{\sqrt{n}}+2^{-n\delta/2}\right) +\frac{|\bar\calX|-3}{2}\log n+c
\\&\ge \inf_{P\in\calP_0} n\Gamma +  \frac{|\bar\calX|-3}{2}\log n+c-\sqrt{nV(P)}\frac{2}{Q'(\eps)}\left(\frac{B+1}{\sqrt{n}}+2^{-n\delta/2}\right)\label{eq:k_bd1}
\\&\ge n\Gamma+  \frac{|\bar\calX|-3}{2}\log n+c-\frac{2|\log\beta|}{Q'(\eps)} \left(B+1+\sqrt{n}2^{-n\delta/2}\right)\label{eq:k_bd2}
\\&\ge n\Gamma+\frac{|\bar\calX|-3}{2}\log n-d'\label{eq:d_prime}
\end{align}
where in \eqref{eq:k_bd1} $Q'(\eps)$ is the derivative of the $Q$ function at $\eps$, and the bound holds for sufficiently large $n$, in \eqref{eq:k_bd2} we have upper bounded the varentropy as $V(P)\le (\log\beta)^2$, and in \eqref{eq:d_prime}  $d'$ is a constant depending only on $\bar\calX$, $\eps$, and $\Gamma$. The above holds only for sufficiently large $n$; call it $n> n_0$ for some $n_0$. We can extend the result for all $n$ by setting $d_{\Gamma}=\max\{d',n_0\Gamma+\frac{|\bar\calX|-3}{2}\log n_0\}$. Using \eqref{eq:d_prime} and the fact that $k\ge 0$ proves \eqref{eq:to_prove} for all $n$.
\end{IEEEproof}

\subsection{Converse for Fixed-to-Variable Prefix Codes}\label{prefix_converse}

\begin{theorem}
For any $\bar{\calX}\subset\calX$, and $\eps>0$, and any sequence of $n$-length prefix codes $\phi_n$,
\beq
\sup_{\substack{P:\calX_P=\bar{\calX}}} \left[R(\phi_n;\eps,P)-H(P)-\sqrt{\frac{V(P)}{n}}Q^{-1}(\eps)\right] \ge \frac{|\bar{\calX}|-1}{2}\frac{\log n}{n}-O\left(\frac{\log\log n}{n}\right).
\eeq
\end{theorem}
\begin{IEEEproof}
We assume all distributions considered in this proof satisfy $\calX_P=\bar\calX$ (i.e., $P(x)=0$ for $x\notin\bar\calX$).

The theorem follows trivially if the sequence of prefix codes $\phi_n$ is such that
\beq\label{eq:too_big}
\sup_{P} \left[R(\phi_n;\eps,P)-J(P)\right]=\omega\left(\frac{\log n}{n}\right).
\eeq
We therefore may assume there is a constant $C$ so that for all $n,\eps,P$
\beq\label{eq:simple_upper_bound}
R(\phi_n;\eps,P)\le J(P)+ C\frac{\log n}{n}.
\eeq

We define a sequence of non-prefix codes $\phi'_n$ as follows: For each $n$, list all $n$-length sequences by their length $\ell(\phi_n(x^n))$, and then map sequences to variable-length bit-strings in this order (breaking ties arbitrarily). Certainly $R(\phi_n;\eps,P)\ge R(\phi'_n;\eps,P)$ for all $P$. For each $\Gamma\in(0,\log|\bar\calX|)$, let $P_\Gamma$ be a distribution in
\beq
\argmax_{P:J(P)=\Gamma} R(\phi'_n;\eps,P).
\eeq
By Theorem~\ref{thm:general_converse},
\beq\label{eq:Gamma}
R(\phi'_n;\eps,P_\Gamma)\ge  \Gamma+\frac{|\bar\calX|-3}{2}\frac{\log n}{n}-\frac{d_\Gamma}{n}.
\eeq
Recall that $d_{\Gamma}$ may depend on $\bar\calX$, $\eps$, and $\Gamma$, but not $n$ or $\phi_n$, and that $d_\Gamma$ is finite for all $\Gamma\in(0,\log|\bar\calX|)$.

Let $k_0,k_1,\ldots,k_I$ be an increasing sequence of integers $k$ for which $k=nR(\phi_n;\eps,P_\Gamma)$ for some $\Gamma$. Define the following for $i=1,\ldots,I$:
\begin{align}
m_i&:=|\{x^n:k_{i-1}<\ell(\phi_n(x^n))\le k_{i}\}|,\label{eq:m_i_def}\\
r_i&:=\frac{1}{n}\log |\{x^n: \ell(\phi_n(x^n))\le k_i\}|.
\end{align}
By the prefix code constraint, there are no more than $2^{k_i}$ sequences with codeword length at most $k_i$. Thus $r_i\le k_i/n$.
 Let $i(\Gamma)$ be the integer such that $nR(\phi_n;\eps,P_\Gamma)=k_{i(\Gamma)}$. By the definition of $\phi'_n$, for any $\Gamma$, $R(\phi'_n;\eps,P_\Gamma)\le r_{i(\Gamma)}$. Thus we have
\beq
R(\phi'_n;\eps,P_\Gamma)\le r_{i(\Gamma)}\le R(\phi_n;\eps,P_\Gamma).
\eeq
Without loss of generality, we consider two values of $\Gamma$: $1/5$ and $4/5$. In particular
\beq\label{eq:r0_upper}
r_0\le r_{i(1/5)}\le R(\phi_n;\eps,P_{1/5})\le \frac{1}{5}+C\frac{\log n}{n}
\eeq
where the last inequality is from \eqref{eq:simple_upper_bound}. Moreover, noting that $\log|\bar\calX|\ge 1>4/5$, we have
\beq\label{eq:rI_lower}
r_I\ge r_{i(4/5)}\ge R(\phi'_n;\eps,P_{4/5})\ge \frac{4}{5}+\frac{|\bar\calX|-3}{2}\frac{\log n}{n}-\frac{d_{4/5}}{n}.
\eeq
Combining \eqref{eq:r0_upper} and \eqref{eq:rI_lower} gives that for sufficiently large $n$, $r_I-r_0\ge 1/2$. Thus there exists $\bari$ such that $r_{\bari}-r_{\bari-1}\ge\frac{1}{2I}$. We set
\beq
\bar\Gamma=\frac{r_{\bari}+r_{\bari-1}}{2}-\frac{|\bar\calX|-3}{2}\frac{\log n}{n}.
\eeq
Thus by \eqref{eq:Gamma}, for sufficiently large $n$, $R(\phi_n';\eps,P_{\bar\Gamma})>r_{\bari}$. Hence $i(\bar\Gamma)>\bari$ so
\begin{align}
R(\phi_n;\eps,P_{\bar\Gamma})
&=\frac{k_{i(\bar\Gamma)}}{n}
\\&\ge \frac{k_{i+1}}{n}
\\&\ge r_{i+1}
\\&=\frac{r_{i+1}+r_{i}}{2}+\frac{r_{i+1}-r_{i}}{2}
\\&\ge\bar\Gamma+\frac{|\bar\calX|-3}{2}\frac{\log n}{n}+\frac{1}{4I}.\label{eq:bound1}
\end{align}

\renewcommand{\balpha}{\boldsymbol{\alpha}}

Now, by Kraft's inequality
\beq\label{eq:kraft}
\sum_i m_i 2^{-k_{i}}\le 1.
\eeq
Note that this is in fact a slight relaxation of Kraft's inequality, since in \eqref{eq:m_i_def} $\ell(\phi_n(x^n))$ may be strictly smaller than $k_{i}$. Recalling that $R(\phi'_n;\eps,P_\Gamma)\le r_{i(\Gamma)}$ and $R(\phi_n;\eps,P_\Gamma)=k_{i(\Gamma)}/n$, we may lower bound the difference between these two rates by
\beq\label{eq:rate_diff}
R(\phi_n;\eps,P_\Gamma)-R(\phi'_n;\eps,P_\Gamma)\ge -\frac{1}{2} \log\left(2^{-k_{i(\Gamma)}+nr_{i(\Gamma)}}\right)
= -\frac{1}{n} \log\left(2^{-k_{i(\Gamma)}}\sum_{j\le i(\Gamma)}m_j\right).
\eeq
Let
\beq
T:=\min_{i} 2^{-k_i} \sum_{j\le i} m_j.
\eeq
Let $i^\star$ be a minimizing $i$ in the above expression. By the definitions of $\{k_i\}$ and $i(\Gamma)$, there is some $\Gamma^\star$ for which $i(\Gamma^\star)=i^\star$. By \eqref{eq:rate_diff},
\beq
R(\phi_n;\eps,P_{\Gamma^\star})-R'(n,\eps,P_{\Gamma^\star})\ge -\frac{1}{n}\log T.
\eeq
Writing $\alpha_i$ for $m_i 2^{-k_i} $, we may upper bound $T$ by the solution to the linear program
\beq
\begin{array}{ll}
\text{maximize} & t\\
\text{subject to} & \displaystyle \sum_{j\le i} 2^{k_j-k_i} \alpha_j\ge t,\text{ for all }i,\\
&\displaystyle \sum_{i} \alpha_i\le 1\\
&\alpha_i\ge 0 \text{ for all }i
\end{array}
\eeq
where the second inequality is derived from Kraft's inequality in \eqref{eq:kraft}. Let $g_i(\balpha)=\sum_{j\le i} 2^{k_j-k_i}\alpha_j$. We claim that there exists an optimal point for the linear program such that the constraint $g_i(\balpha)\ge t$ holds with equality for all $i$. Suppose not: that $g_i(\balpha)>t$ for some $i$. Then we may form a different feasible point with the same value of $t$ as follows. Let $i_1$ be the smallest $i$ such that $g_i(\balpha)>t$. Let $\Delta>0$ be such that $g_{i_1}(\balpha)=t+\Delta$. Note that for all $i>0$
\beq\label{eq:g_recursion}
g_i(\balpha)=\alpha_i+2^{k_{i-1}-k_i}g_{i-1}(\balpha).
\eeq
For convenience, we adopt the convention that $k_{-1}=-\infty$ and $g_{-1}(\balpha)=t$. With this convention \eqref{eq:g_recursion} holds even for $i=0$. Moreover, since by assumption $g_{i_1-1}(\balpha)=t$, we have that
\beq
g_{i_1}(\balpha)=t+\Delta=\alpha_{i_1}+2^{k_{i_1-1}-k_{i_1}}t.
\eeq
Thus $\alpha_{i_1}=(1-2^{k_{i_1-1}-k_{i_1}})t+\Delta \ge \Delta$. We define a vector $\balpha'$ where $\alpha'_i=\alpha_i$ except that $\alpha'_{i_1}=\alpha_{i_1}-\Delta$ and $\alpha'_{i_1+1}=\alpha_{i_1}+\Delta$, where if $i_1=I$ then the latter does not apply. Note that $\alpha'_i\ge 0$ for all $i$, and that $\sum_i \alpha'_i\le \sum_i\alpha_i\le 1$. Moreover, for all $i<i_1$ we have $g_i(\balpha')=g_i(\balpha)=t$. Also by construction $g_{i_1}(\balpha')=t$. For $i>i_1$
\beq
g_i(\balpha')-g_i(\balpha)=2^{k_{i_1}-k_i}(\alpha'_{i_1}-\alpha_{i_1})+2^{k_{i_1+1}-k_i}(\alpha'_{i_1+1}-\alpha_{i_1+1})
= -2^{k_{i_1}-k_i}\Delta+2^{k_{i_1+1}-k_i}\Delta\ge 0.
\eeq
Hence for $i>i_1$ we have $g_i(\balpha')\ge g_i(\balpha)\ge t$. Thus $\balpha'$ is a feasible point, with an additional equality $g_{i_1}(\balpha')=t$ that did not hold for $\balpha$. Repeating this procedure yields a feasible point for the same $t$ where $g_i(\balpha)=t$ for all $i$. Therefore, there exists an optimal point for the linear program that satisfies these equalities. For this optimal point, by \eqref{eq:g_recursion}, for all $i$, $t=\alpha_i+2^{k_{i-1}-k_i}t$, so $\alpha_i=(1-2^{k_{i-1}-k_i})t$. From the constraint that $\sum_i \alpha_i\le 1$, we have
\beq
\sum_{i} (1-2^{k_{i-1}-k_i})t\le 1.
\eeq
Therefore the optimal value for $t$ is
\beq
t=\frac{1}{\sum_{i} (1-2^{k_{i-1}-k_i})}.
\eeq
Thus
\begin{equation}
R(\phi_n;\eps,P_{\Gamma^\star})-R(\phi'_n;\eps,P_{\Gamma^\star})
\ge \frac{1}{n}\log\left(\sum_{i} (1-2^{k_{i-1}-k_i})\right)
\ge \frac{1}{n}\log \frac{I}{2}
\end{equation}
where we have used the fact that $k_i-k_{i-1}\ge 1$. Therefore by \eqref{eq:Gamma}
\beq\label{eq:bound2}
R(\phi_n;\eps,P_{\Gamma^\star})\ge \Gamma^\star+\frac{|\calX|-3}{2}\frac{\log n}{n}+\frac{1}{n}\log \frac{I}{2}- \frac{d_{\Gamma^\star}}{n}.
\eeq

Combining \eqref{eq:bound1} with \eqref{eq:bound2}, for sufficiently large $n$ 
\beq
\sup_P \big[R(\phi_n;\eps,P)-J(P)\big]
\ge \frac{|\bar\calX|-3}{2}\frac{\log n}{n} +\max\left\{\frac{1}{4I},\frac{1}{n}\log \frac{I}{2}\right\}-\frac{d_{\Gamma^\star}}{n}.
\eeq
Since $\frac{1}{4I}$ is decreasing in $I$ and $\frac{1}{n}\log \frac{I}{2}$ is increasing in $I$, for any $I'$
\beq
\min\left\{\frac{1}{4I'},\frac{1}{n}\log \frac{I'}{2}\right\}\le \inf_I \max\left\{\frac{1}{4I},\frac{1}{n}\log \frac{I}{2}\right\}\le \max \left\{\frac{1}{4I'},\frac{1}{n}\log \frac{I'}{2}\right\}.
\eeq
Thus, if we choose $I'=\frac{n}{4(\log n-\log\log n)}$ we find
\beq
\inf_{I} \max\left\{\frac{1}{4I},\frac{1}{n}\log \frac{I}{2}\right\}=\frac{\log n}{n}-O\left(\frac{\log \log n}{n}\right).
\eeq
Therefore
\beq
\sup_P \left[R(\phi_n;\eps,P)-J(P)\right]
\ge \frac{|\bar\calX|-1}{2}\frac{\log n}{n} -O\left(\frac{\log\log n}{n}\right).
\eeq
\end{IEEEproof}

\section{Concluding Remarks}

\label{SecConc}

We have derived achievability and converse bounds on the third-order coding rates for universal prefix-free and prefix fixed-to-variable codes. This required the new Type Size code for prefix-free achievability, unlike traditional Two-Stage codes. The converse involved an approach based on mixture distributions, bounds on the empirical entropy, and Laplace's approximation. Future work includes studying sources with memory and lossy coding.

\bibliographystyle{IEEEtran}
\bibliography{KS_Proposal}

\end{document}